\documentstyle[12pt,fleqn]{article}
\topmargin - 0.8cm

\renewcommand{\theequation}{\arabic{section}.\arabic{equation}}

\newcommand{\I}{\mbox{i}}
\newcommand{\E}{\mbox{e}}
\newcommand{\D}{\mbox{d}}
\catcode`@=11
\@addtoreset{equation}{section}
\catcode`@=12
\mathchardef\SGamma="7100

\begin{document}
\begin{titlepage}
\begin{flushright}
Freiburg THEP-97/25\\
gr-qc/9711037
\end{flushright}
\vskip0.7cm
\begin{center}
\vfill
{\large \bf Wheeler-DeWitt equation
and Feynman diagrams}
\vskip1cm
{\bf Andrei O. Barvinsky}
\vskip0.4cm
Theory Department, Lebedev Physics Institute and Lebedev Research Center in
Physics,\\ Leninsky Prospect 53,
Moscow 117924, Russia.
\vskip0.6cm
{\bf Claus Kiefer}
\vskip0.4cm
Faculty of Physics, University of Freiburg,
Hermann-Herder-Stra\ss e 3,\\ D-79104 Freiburg, Germany.
\end{center}
\vskip2cm
\begin{center}
{\bf Abstract}
\end{center}
\begin{quote}
We present a systematic expansion of all constraint
equations in canonical quantum gravity up to the order of the inverse
Planck mass squared. It is demonstrated that this method generates
the conventional Feynman diagrammatic technique involving graviton loops
and vertices. It also reveals explicitly
 the back reaction effects of quantized matter
and graviton vacuum polarization. This provides an explicit correspondence
between the frameworks of canonical and covariant quantum gravity
in the semiclassical limit.
\end{quote}
\vskip3cm
\begin{center}
{\em Submitted to Nuclear Physics B}
\end{center}
\end{titlepage}

\section{Introduction}
\hspace{\parindent}
Attempts to understand the quantum nature of the gravitational field
typically fall into two classes. One possibility is to first construct
a well-defined mathematical framework for the full theory,
from which eventually all quantum gravitational phenomena can be
derived. Superstring theory and canonical quantum gravity in the
Ashtekar formulation belong to this class. The other alternative
seeks to apply various approximation schemes to a set of formal
quantum gravity equations. One hopes that in this way some of the
profound conceptual problems of quantum gravity and quantum cosmology
-- such as the problem of time and the related problem of Hilbert
space -- can be tackled. One also hopes to derive in this way
quantum gravitational corrections to standard physics.
Although thus in a sense less ambitious than the first class
of alternatives, it is far more straightforward to get definite
results which even may provide the first window towards genuine tests
of quantum gravity.

One such approximation scheme is to perform an expansion of
the Wheeler-DeWitt equation of canonical quantum gravity with respect
to the inverse Planck mass squared, $m_P^{-2}$. This expansion
was pioneered in \cite{DW,RL} and later developed from various
points of view. A most remarkable result is the recovery
-- at order $m_P^0$ -- of the (functional) Schr\"odinger equation
on a classical background spacetime as an {\em approximate}
equation from quantum gravity \cite{Banks,HHal,Ha,Ki87}.
At this order, however, two important aspects of the full theory
are not yet incorporated:
Back reaction effects of quantum
matter on the dynamical gravitational background and the proper quantum
effects of the gravitational field itself. Their description requires
higher-order iterations of the Wheeler-DeWitt equations in inverse powers of
the Planck mass.
Such higher-order correction terms were derived in various forms
in both the canonical theory \cite{barv89,Kiefer1,Kiefer2}
and in the covariant framework of path integrals
\cite{barvin,GenSem,BarvU}.

In \cite{Kiefer1}, definite correction terms to the Schr\"odinger
equation were derived at order $m_P^{-2}$ from the Wheeler-DeWitt
equation in a formal way (without addressing the issue of
regularization). In particular cases it was possible to
evaluate these correction terms to find quantum gravitationally
induced energy shifts \cite{Kiefer2,Kiefer}. It was also shown that
these correction terms can in principle lead to observable effects
in the spectrum of the cosmological background radiation \cite{Ro}.

Apart form the issue of regularization, the analysis in \cite{Kiefer1}
was incomplete for two reasons: First, it applied
the expansion scheme only to the quantum Hamiltonian constraint, but not
to the full, complete, set of constraint equations. A complete
picture has to implement all constraints which are interconnected
by their commutator algebra \cite{Kiefer3}.
Second, although a general expression for all correction terms
to the Schr\"odinger equation was given in \cite{Kiefer1}, the final
expression of these terms only involved the so-called ``longitudinal"
derivatives -- the derivatives with respect to configuration space
coordinates along the classical trajectory.
Only this part of the correction terms follows from the previous order
equations alone, while the remaining terms encode the dependence
on boundary conditions. The main point of the present paper is to
present a concise method of giving explicit expressions for
{\em all} correction terms (including the ``transversal
derivatives") for the {\em full} set of quantum constraints.
It is shown that this leads to the conventional
Feynman diagrammatic technique involving graviton propagator, vertices and
loops and thus provides a concrete physical interpretation
of all terms.

One of the widespread approaches to canonical quantum gravity is
characterized by the fact that it
does not contain a notion of spacetime
at the fundamental level (see e.g. \cite{Ringberg} for a conceptual
discussion). Spacetime emerges as an {\em approximate}
notion on the level of the semiclassical expansion described above.
But even within such a conceptual framework it would be desirable
on this level to possess a spacetime
{\em covariant} description of all correction terms. This
would yield a bridge between the canonical formalism and the covariant
effective action formalism \cite{barvin,DW:Dynamical}.
It would be
of crucial importance for the correct covariant regularization of the
ultraviolet divergences which inevitably
arise in all loop orders of the $1/m_P^2$-expansion. This
regularization should maintain the general covariance in the form not
splitted by a (3+1)-foliation. The general covariance is not manifest due to
the canonical origin of the constraint equations, but it gets restored
in the proposed calculation technique, because the arising Feynman diagrams
can be cast into spacetime covariant form. Another important advantage
of this technique is that it clearly reveals the back reaction effects of
quantum matter and graviton vacuum polarization. This back reaction
has the form of special nonlocal gravitational potentials (similar to
the retarded potentials in electrodynamics) contributing to the kinetic
and potential terms in the effective Hamiltonian of the
corrected Schr\"odinger equation.

Our paper is organized as follows. Section~2 gives a brief introduction
into the formalism of the Planck mass expansion. In Section~3
we review the work in \cite{BKr,Barv} presenting a consistent
factor ordering of the theory. We then give the calculation
of the one-loop gravitational prefactor.
In Section~4 we present a detailed calculation of all correction
terms to the matter Hamiltonian up to order $m_P^{-2}$.
As we shall describe, there will be three types of correction terms:
a purely quantum gravitational term, a term corresponding to the
contribution of quantum matter, and a cross term.
Sections~5--7 give a description of all the correction terms
in the language of Feynman diagrams (some computational details
are relegated to appendices).
Section~8 gives an account of the back reaction of matter on the
gravitational background as well as a comparison with the
terms calculated in \cite{Kiefer1}.
Section~9 contains a summary and an outlook on future work.

\section{The Wheeler-DeWitt and Schr\"odinger equations}
\hspace{\parindent}
Consider the Lagrangian action of Einstein gravity theory coupled to matter,
        \begin{eqnarray}
        S^{\rm tot}[g_{\alpha\beta}(x),\varphi(x)]=m_P^2\int \D^4x
        \sqrt{g}\,{}^4\!R(g_{\alpha\beta}(x))
        +S^{\rm mat}[g_{\alpha\beta}(x),\varphi(x)]\ ,
        \end{eqnarray}
where $m_P^2\equiv 1/16\pi G$, and the cosmological constant
is set equal to zero.
The canonical form of this action in terms of three-metric coefficients $g_{ab}$,
matter fields $\varphi$ and their conjugate momenta
$p^{ab}, p_{\varphi}$ ($a,b,...=1,2,3$) has the form \cite{DW,ADM,KucharH}
        \begin{eqnarray}
        S^{\rm tot}[g_{ab}, p^{ab},\varphi,p_{\varphi},N^{\perp},
         N^a]=
        \int \D t\,\D^3x\,\left( m_P^2\, p^{ab}\dot{g}_{ab}+
        p_{\varphi}\dot{\varphi}
        -N^{\perp}T_{\perp}-N^{a}T_a\right),       \label{2.1}
        \end{eqnarray}
where the lapse $N^{\perp}=(-{}^4\!g^{00})^{-1/2}$ and shift
$N^a=g^{ab}{}^4\!g_{0b}$ functions enter as Lagrange multipliers of the
gravitational constraints
        \begin{eqnarray}
        &&T_\perp=m_P^2\,H_{\perp}(g_{ab},\varphi,p^{ab},p_{\varphi})+
        H_\perp^{\rm mat}
        (g_{ab},\varphi,p^{ab},p_{\varphi}),               \label{2.1a}\\
        &&T_a=m_P^2\,H_a(g_{ab},\varphi,p^{ab},p_{\varphi})+
        H_a^{\rm mat} (g_{ab},\varphi,p^{ab},p_{\varphi}).     \label{2.1b}
        \end{eqnarray}
The gravitational parts of these constraints are given by the gravitational
superhamiltonian and supermomentum
        \begin{eqnarray}
        &&H_{\perp}(g_{ab},\varphi,p^{ab},p_{\varphi})=
        \frac 12 G_{ab,cd}p^{ab}p^{cd}-
        \sqrt{g}\,{}^3\!R,                      \label{2.2}\\
        &&H_a(g_{ab},\varphi,p^{ab},p_{\varphi})=
        -2g_{ab}\nabla_c p^{bc},                 \label{2.3}
        \end{eqnarray}
where $G_{ab,cd}=g^{-1/2}(g_{ac}g_{bd}+g_{ad}g_{bc}-g_{ab}g_{cd})$ is a
local DeWitt supermetric and $\nabla_c$ denotes the covariant spatial
derivative. The matter parts of the constraints, $H_{\perp}^{mat}$
 and $H_a^{mat}$,
depend on the concrete choice of matter action
which we shall not specify here. Its form can be strongly constrained
from general principles such as ultralocality \cite{Tei}.

The above constraints
        \begin{eqnarray}
        T_{\perp}\approx0,\quad T_a\approx 0,   \label{2.4}
        \end{eqnarray}
are in Dirac's terminology first class constraints
\cite{Dirac} having
 the following closed Poisson algebra \cite{DW,ADM,KucharH,Tei}
        \begin{eqnarray}
        &&\{T_\perp({\bf x}),T_\perp({\bf x}')\}=
        g^{ab}({\bf x})T_b({\bf x})
        \partial_a\delta({\bf x},{\bf x}')-
        ({\bf x}\leftrightarrow{\bf x}'),            \nonumber\\
        &&\{T_\perp({\bf x}),T_a({\bf x}')\}=
        -T_\perp({\bf x}')\,
        \partial_a\delta({\bf x}',{\bf x}),            \nonumber\\
        &&\{T_a({\bf x}),T_b({\bf x}')\}=T_b({\bf x})
        \partial_a\delta({\bf x},{\bf x}')-
        (a,{\bf x}\leftrightarrow b,{\bf x}')\ .            \label{2.4a}
        \end{eqnarray}

These constraints imply that the phase space variables are not dynamically
independent -- a property which in the Dirac quantization framework
leads to quantum constraints on physically admissible
quantum states. In this quantization scheme the c-number constraints
(\ref{2.1a}) -- (\ref{2.1b}) become the operators
$(\hat{T}_{\perp},\hat{T}_a)$ which should satisfy
the closed commutator algebra generalizing (\ref{2.4a}) to the quantum
domain. They select the physical quantum states
$|\,{\mbox{\boldmath$\Psi$}}\big>$ by the equations \cite{DW,Dirac}
        \begin{eqnarray}
        \hat{T}_{\perp}|{\mbox{\boldmath$\Psi$}}\big>=0,\quad
        \hat{T}_a\,|{\mbox{\boldmath$\Psi$}}\big>=0. \label{2.8}
        \end{eqnarray}

In what follows we shall denote the quantum states in the standard
 representation (Hilbert) space of matter fields
$(\hat{\varphi}({\bf x}),\hat{p}_\varphi({\bf x}))$ by ket-vectors,
the matter field operators label by hats and choose the functional
coordinate representation for metric variables
        \begin{equation}
        \hat{g}_{ab}({\bf x})=g_{ab}({\bf x}), \quad
        \hat{p}^{ab}({\bf x})=\frac 1{\I m_P^2}
        \frac{\delta}{\delta g_{ab}({\bf x})}.             \label{2.9}
        \end{equation}
Note that we have redefined the gravitational momenta by a factor
$m_P^2$ compared with the standard convention.

The quantum states then become the state vector-valued functionals
of three-metric coefficients
$|{\mbox{\boldmath$\Psi$}}[g_{ab}({\bf x})]\big>$, and the equations
(\ref{2.8}) take the form of well known Wheeler-DeWitt equations
        \begin{eqnarray}
        &&\mbox{\vphantom{$\left\{\frac{L^{L^{L^a}}_{L}}
        {L^L_L g}\right\}$}}^{\prime\prime}\!
        \left\{-\frac 1{2 m_P^2}
        G_{ab,cd}\frac{\delta^2}
        {\delta g_{ab}\delta g_{cd}}-m_P^2\,\sqrt{g}\,{}^3\!R +
        \hat{H}{}_{\perp}^{mat}\right\}^{\prime\prime}
        |{\mbox{\boldmath$\Psi$}}[g_{ab}]\big>=0,           \label{2.10}
        \\
        &&\mbox{\vphantom{$\left\{\frac{L^{L^{L^a}}_{L}}
        {L^L_L g}\right\}$}}^{\prime\prime}\!
        \left\{-\frac{2}\I g_{ab}\nabla_c
        \frac{\delta}{\delta g_{bc}}+\hat{H}{}_a^{mat}
        \right\}^{\prime\prime}
        |{\mbox{\boldmath$\Psi$}}[g_{ab}]\big>=0,            \label{2.11}
        \end{eqnarray}
where the inverted commas indicate that these functional variational operators
are a symbolic representation of some operator realization of the classical
constraints (\ref{2.2}) -- (\ref{2.3}), implying both the operator ordering
and quantum corrections proportional to $\hbar$.

The form of (\ref{2.10}) and (\ref{2.11}) shows that the size
of the quantum gravitational effects is governed by the dimensional parameter
$m_P^2$ - the square of the Planck mass. The quantum effects of matter fields
(their quantum commutators, couplings, etc.) in turn are determined by
$\hbar$ chosen for simplicity to be one in our units. The justification
of this choice is that in our paper we shall develop the semiclassical
expansion in $1/m_P^2$ of quantum gravitational effects without
$\hbar$-expansion of quantum matter. The lowest-order approximation of such
an expansion for solutions of the Wheeler-DeWitt equations
\cite{DW,RL,Banks,Kiefer2} yields
      the following semiclassical form
        \begin{eqnarray}
        |{\mbox{\boldmath$\Psi$}}[g_{ab}]\big>=
        \E^{\,{}^{\textstyle{\I m_P^2
        {\mbox{\boldmath$S$}[g_{ab}]}}}}|\Phi [g_{ab}]\big>\ , \label{2.12}
        \end{eqnarray}
where ${\mbox{\boldmath$S$}[g_{ab}]}$ is a purely gravitational
Hamilton-Jacobi function. This is a solution of the vacuum
Einstein-Hamilton-Jacobi equations -- the gravitational constraint equations
with the momenta replaced by their Hamilton-Jacobi values (gradients
of ${\mbox{\boldmath$S$}[g_{ab}]}$):
        \begin{eqnarray}
        &&H_{\perp}(g_{ab},\delta{\mbox{\boldmath$S$}}/\delta g_{ab})
        \vphantom{\frac11}\equiv\frac 12 G_{ab,cd}
        \frac{\delta{\mbox{\boldmath$S$}}}
        {\delta g_{ab}}
        \frac{\delta{\mbox{\boldmath$S$}}}{\delta g_{cd}}
        -\sqrt{g} ^3\!R=0,                             \label{2.13}\\
        &&H_a(g_{ab},\delta{\mbox{\boldmath$S$}}/\delta g_{ab})
        \equiv-2g_{ab}\nabla_c
        \frac{\delta{\mbox{\boldmath$S$}}}
        {\delta g_{bc}}=0 \ .                             \label{2.14}
        \end{eqnarray}

Substitution of (\ref{2.12}) into the Wheeler-DeWitt equations leads
to new equations for the state vector of matter fields $|\Phi[g_{ab}]\big>$
parametrically depending on the spatial metric
        \begin{eqnarray}
        &&\left \{\frac 1\I G_{ab,cd}
        \frac{\delta{\mbox{\boldmath$S$}}}
        {\delta g_{ab}} \frac{\delta}
        {\delta g_{cd}}+\hat{H}_{\perp}^{mat}(g_{ab})\right.\nonumber\\
        &&\qquad\quad\quad\quad\quad\quad
        \left.+\frac 1{2\I}\,
        \mbox{\vphantom{$\left\{\frac{L^{L^{L^a}}_{L}}
        {L^L_L g}\right\}$}}^{\prime\prime}\!\!
        G_{ab,cd}
        \frac{\delta^2{\mbox{\boldmath$S$}}}
        {\delta g_{ab}\delta g_{cd}}^{\prime\prime}
        -\frac 1{2m_P^2}\,\mbox{\vphantom{$\left\{\frac{L^{L^{L^a}}_{L}}
        {L^L_L g}\right\}$}}^{\prime\prime}\!\!
        G_{ab,cd}\frac{\delta^2}
        {\delta g_{ab}\delta g_{cd}}^{\prime\prime}
        \right\}
        |\Phi[g_{ab}]\big>=0,                  \label{2.15}\\
        &&\left\{\mbox{\vphantom{$\left\{\frac{L^{L^{L^a}}_{L}}
        {L^L_L g}\right\}$}}^{\prime\prime}\!\!
        -\frac{2}\I g_{ab}\nabla_c
        \frac{\delta}{\delta g_{bc}}^{\prime\prime}+
        \hat{H}_a^{mat}(g_{ab})\right\}|\Phi[g_{ab}]\big>=0. \label{2.16}
        \end{eqnarray}

The conventional derivation of the Schr\"odinger equation from the
Wheeler-DeWitt equations then consists in the assumption of {\it small
back reaction of quantum matter on the metric background} which
at least heuristically allows one to discard the third and the fourth
terms in (\ref{2.15}). Then one considers $|\Phi[g_{ab}]\big>$ on the
solution of classical vacuum Einstein equations $g_{ab}({\bf x},t)$
corresponding to the Hamilton-Jacobi function
 ${\mbox{\boldmath$S$}[g_{ab}]}$,
        \begin{eqnarray}
        |\Phi(t)\big>=|\Phi[g_{ab}({\bf x},t)]\big>. \label{2.17}
        \end{eqnarray}
After a certain {\em choice} of lapse and shift functions $(N^{\perp},N^a)$,
this solution satisfies the canonical equations
        \begin{eqnarray}
        \dot{g}_{ab}=N^{\perp}G_{ab,cd}
        \frac{\delta{\mbox{\boldmath$S$}}}{\delta g_{cd}}+
        2\nabla_{(a}N_{b)},               \label{2.18}
        \end{eqnarray}
so that the quantum state (\ref{2.17}) satisfies the evolutionary equation
obtained by using
        \begin{eqnarray}
        \frac{\partial}{\partial t}\,|\Phi(t)\big>=
        \int \D^3 x \,\dot{g}_{ab}({\bf x})\,
        \frac{\delta}{\delta g_{ab}({\bf x})}
        |\Phi[g_{ab}]\big>
        \end{eqnarray}
together with the truncated version of equations (\ref{2.15}) -- (\ref{2.16}).

The result in the approximation of the above type is
the Schr\"odinger equation of quantized matter fields in the external classical
gravitational field,
        \begin{eqnarray}
        &&\I\frac{\partial}{\partial t}\,
        |\Phi(t)\big>=\hat{H}{}^{\rm mat}|\Phi(t)\big>, \\
        &&\hat{H}{}^{\rm mat}=
        \int \D^3 x \left\{N^{\perp}({\bf x})
        \hat{H}{}^{\rm mat}_{\perp}({\bf x})+
        N^a({\bf x})\hat{H}{}^{\rm mat}_a({\bf x})\right\}.     \label{2.19}
        \end{eqnarray}
Here, $\hat{H}{}^{\rm mat}$ is a matter field Hamiltonian in the Schr\"odinger
picture, parametrically depending on (generally nonstatic) metric
coefficients of the curved spacetime background.
Examples can be found e.g. in \cite{Kiefer2,Kiefer}.

Such a derivation of quantum field theory from the Wheeler-DeWitt equations
dates back, at the level of cosmological models, to the pioneering work of
DeWitt \cite{DW}. It was later performed by Lapchinsky and Rubakov
\cite{RL} for generic gravitational systems and discussed in various contexts
in \cite{Banks,HHal,Ha,Ki87,barv89,Kiefer1,Kiefer2}. This method turns out
to be the most
commonly used approach and, in fact, the most effective way of interpreting
the cosmological wavefunction in the semiclassical approximation.
 It establishes in particular the links between fundamental
quantum cosmology and the physics of the early Universe. On the basis of the
obtained
Schr\"odinger equation this method obviously generates all quantum effects of
matter fields in curved spacetime, but in this
order of approximation it does not yet
contain two important ingredients of the complete quantum scheme: i) purely
gravitational quantum effects and ii) back reaction effects of quantum matter
on the gravitational fields. Obviously, these effects are generated by the
discarded third and fourth terms of (\ref{2.15}). The purpose of this paper
is to develop the systematic method of $1/m_P^2$-expansion perturbatively
generated by these terms and demonstrate that this method gives rise to
back reaction effects and can be described by a
Feynman diagrammatic technique including graviton
and graviton-matter loops.

{}From a methodological point of view, we shall show
 how the treatment of the gravitational preexponential factor
presented in \cite{GenSem,BarvU,BKr,Barv}
improves the results of \cite{barv89} where the purely
gravitational effects were not properly disentangled. We
shall also give transparent expression for the ``transversal
derivatives" of the correction terms \cite{Kiefer1}.
As we shall
see, these terms generate the kinetic part
 of the nonlocal gravitational potentials
describing the back reaction of quantum matter on the gravitational
background.

There is an obvious problem arising in the implementation of this program.
The effects we are going to consider go beyond the tree-level approximation
and, therefore, require the precise knowledge of the operator realization
of the Wheeler-DeWitt equations (\ref{2.10}) -- (\ref{2.11}) and
(\ref{2.15}) -- (\ref{2.16}), that is replacing the inverted commas by actual
operator ordering of coordinates and momenta. The main property that these
operators should satisfy is a closure of the quantum commutator algebra
of constraints generalizing (\ref{2.4a}) to the quantum level as consistency
conditions of (\ref{2.8}). In the coordinate representation this
realization is known for {\em generic} theories subject to first-class
constraints in the {\em one-loop} approximation \cite{BKr}. Fortunately,
for theories with constraints of the {\em gravitational type} quadratic
and linear in momenta (like (\ref{2.2}) -- (\ref{2.3})) this realization is
known in the exact theory, formally closing the commutator algebra beyond
the semiclassical expansion \cite{Barv}\footnote{
The formal closure of the commutator algebras of \cite{BKr,Barv}
holds up to possible quantum anomalies arising when regulating ill-defined
products of operators taken at the same space point. The problem of these
anomalies in field models in spacetime dimensions higher than two is far
from resolution and goes beyond the present paper.
}. In the next section we dwell on the construction of this realization in
terms of the condensed DeWitt notation which will also allow us to
formulate the necessary gravitational prefactor before going over to
higher orders of the $1/m_P^2$-expansion for the Wheeler-DeWitt equations.

\section{Operator realization of the Wheeler-DeWitt equations
and the one-loop gravitational prefactor}
\hspace{\parindent}
The operator realization of the Wheeler-DeWitt equations in the full
theory not restricted by any minisuperspace reductions uses a very
complicated formalism. What makes this formalism manageable is the use of
condensed DeWitt notations which formally represents the complicated
field system in terms of a quantum-mechanical model with a finite-dimensional
phase space having a finite-dimensional
space of local gauge (diffeomorphism) transformations. This
can be achieved by introducing the collective notation for gravitational
phase space variables
        \begin{eqnarray}
        q^i=g_{ab}({\bf x}),\,\,\,
        p_i=p^{ab}({\bf x}),         \label{31}
        \end{eqnarray}
in which the condensed index $i=(ab, {\bf x})$ includes both discrete
tensor indices and
three-dimensional spatial coordinates ${\bf x}$. Similar notations for
the full set of constraints and their gravitational and matter parts yields
        \begin{eqnarray}
        &&T_{\mu}(q,p)=(T_{\perp}({\bf x}),T_a({\bf x})),\quad
        T_{\mu}(q,p)=m_P^2\,H_{\mu}+H_{\mu}^{\rm mat},         \label{32}\\
        &&H_{\mu}(q,p)=(H_{\perp}({\bf x}),H_a({\bf x})), \quad
        H_{\mu}^{\rm mat}(q,\varphi,p_\varphi)=
        (H^{\rm mat}_{\perp}({\bf x}),H^{\rm mat}_a({\bf x}))\ .   \label{33}
        \end{eqnarray}
The index $\mu$ enumerates the superhamiltonian and
supermomenta of the theory, as well as their spatial coordinates $\mu\to
(\mu,{\bf x})$. In these notations the functional dependence on phase space
variables is represented in the form of functions of $(q^i,p_i)$, and
the contraction of condensed indices includes integration over ${\bf x}$
along with discrete summation. In condensed notations the gravitational
part of the canonical action (\ref{2.1}) has the simple form
        \begin{eqnarray}
        S[q,p,N]=\int \D t\, \left( p_i\dot{q}^i-
        N^{\mu}H_{\mu}(q,p)\right),                       \label{34}
        \end{eqnarray}
with the superhamiltonian and supermomenta given by expressions
which are quadratic and linear in the momenta, respectively,
        \begin{eqnarray}
        H_{\perp} =\frac 12 G^{\,ik}_{\perp} p_i p_k +V_\perp, \quad
        H_a=\nabla^i_a p_i.                         \label{35}
        \end{eqnarray}
Here the indices $\perp\to(\perp,{\bf x})$ and $a\to(a,{\bf x})$
are also condensed, $G^{ik}_{\perp}$ is the ultralocal three-point
object containing the matrix of the DeWitt supermetric, $V_\perp$ denotes
the potential term of the superhamiltonian (\ref{2.2}), and $\nabla_a^i$ is
the generator of the spatial diffeomorphism of $q^i$ (see below). The objects
$G^{ik}_{\perp}$ and $\nabla_a^i$ have the form of the following
delta-function type kernels \cite{Barv}
        \begin{eqnarray}
        &&G^{\,ik}_{\perp}=G_{ab,cd}\ \delta({\bf x}_i,{\bf x}_k)
        \delta({\bf x}_\perp,{\bf x}_k),\,\,\, i=(ab,{\bf x}_i),\,\,\,
        k=(cd,{\bf x}_k),\,\,\,
        \perp=(\perp,{\bf x}_\perp),        \label{36}\\
        &&\nabla_a^i=-2g_{a(b}\nabla_{c)}
        \delta({\bf x}_a,{\bf x}_i),\,\,\,
        i=(bc,{\bf x}_i),\,\,\, a=(a,{\bf x}_a). \label{37}
        \end{eqnarray}

Note that the object $G^{\,ik}_{\perp}$ itself does not form the DeWitt
supermetric because it contains two delta-functions. Only the functional
contraction of $G^{\,ik}_{\perp}$ with the constant lapse function
$N^{\perp}=1$ converts this object into the distinguished ultralocal
metric on the functional space of three-metric coefficients,
        \begin{eqnarray}
        G^{\,ik}=\left.G^{\,ik}_{\perp}N^{\perp}
        \right|_{N^{\perp}=1}
        \equiv \int\D^3{\bf x}_{\perp}\ G^{\,ik}_{\perp}
        =G_{ab,cd}\ \delta({\bf x}_i,{\bf x}_k).           \label{38}
        \end{eqnarray}

The Poisson bracket algebra for the gravitational constraints
in condensed notations can be written as
        \begin{eqnarray}
        \{H_\mu,H_\nu\}=U^{\alpha}_{\mu\nu}H_\alpha       \label{39}
        \end{eqnarray}
with structure functions $U^{\alpha}_{\mu\nu}=U^{\alpha}_{\mu\nu}(q)$ that
can be read off from (\ref{2.4a}). This algebra implies the gauge invariance
of the action (\ref{34}) under the transformations with local (arbitrary
time and space dependent) parameters ${\cal F^\mu}={\cal F^\mu}(t)$. These
transformations are canonical (ultralocal in time) for phase space variables,
        \begin{eqnarray}
        \delta q^i=\{q^i,H_\mu\}{\cal F^\mu},\,\,\,
        \delta p_i=\{p_i,H_\mu\}{\cal F^\mu},           \label{40}
        \end{eqnarray}
and quasilocal (involving the time derivative of ${\cal F^\mu}$) for
Lagrange multipliers \cite{BFV},
        \begin{eqnarray}
        \delta N^\mu=\dot{\cal F^\mu}
        -U^{\mu}_{\alpha\nu}N^\alpha{\cal F^\nu}.        \label{41}
        \end{eqnarray}
Note that the transformations of phase space {\em coordinates}
        \begin{eqnarray}
        \delta q^i=\nabla^i_\mu{\cal F^\mu},\quad
        \nabla^i_\mu\equiv
        \frac{\partial H_{\mu}}{\partial p_i}               \label{42}
        \end{eqnarray}
have as generators the vectors $\nabla^i_\mu$ which are momentum
independent for spacelike diffeomorphisms $\mu=a$ (and, therefore,
coincide with the coefficients of the momenta in the supermomentum constraints
(\ref{37})), but involve momenta for spacetime diffeomorphisms normal
to spatial slices, $\nabla^i_\perp=G^{\,ik}_{\perp}p_k$.

With these condensed notations let us formulate the operator
realization of gravitational constraints
$H_\mu(q,p)\rightarrow\hat{H}_{\mu}$ closing the commutator
version of the Poisson bracket algebra (\ref{2.4a})
        \begin{eqnarray}
        \left[\hat{H}_\mu,\hat{H}_\nu\right]=
        \I\hat{U}^{\lambda}_{\mu\nu}\hat{H}_\lambda.  \label{43}
        \end{eqnarray}
As shown in \cite{Barv}, the fact that (\ref{43}) holds
 follows from the classical gravitational
constraints (\ref{35}) by replacing the momenta $p_k$ with the functional
covariant derivatives ${\cal D}_k/im_P^2$ -- covariant with respect to the
Riemann connection based on the DeWitt supermetric (\ref{38}) -- and by
adding a purely imaginary part (anti-Hermitian with respect to the $L^2$ inner
product): the functional trace of structure functions,
$\I\hbar U^{\nu}_{\mu\nu}/2$. With this definition of covariant derivatives
it is understood that the superspace of three-metrics $q$ is regarded as
a functional differentiable manifold, and quantum states
$|{\mbox{\boldmath$\Psi$}}(q)\Big>$ are scalar densities of 1/2-weight.
Thus the operator realization for the full constraints including the matter
parts has the form \cite{Barv}\footnote
{
We assume that the operators of matter parts of constraints
$\hat{H}{}^{\rm mat}_\mu$ independently satisfy the commutator algebra
with the {\em same} operator structure functions as in the gravitational
sector (\ref{39}), while the commutator between $\hat{H}_\mu$ and
$\hat{H}{}^{\rm mat}_\mu$ is identically zero \cite{Tei}. With
respect to possible quantum anomalies this assumption is rather nontrivial,
but this goes, as it has already been mentioned above, beyond the scope of
this paper.
}
        \begin{eqnarray}
        &&\hat{T}_{\perp} =-\frac 1{2m_P^2} G^{\,ik}_{\perp}
        {\cal D} _i {\cal D}_k +m_P^2\,V_\perp+
        \frac{\I}2 U^{\nu}_{\perp\nu}
        +\hat{H}{}^{\rm mat}_\perp,              \label{44}\\
        &&\hat{T}_a=\frac 1\I \nabla^i_a {\cal D}_i+
        \frac{\I}2 U^{\nu}_{a\nu}
        +\hat{H}{}^{\rm mat}_a.                    \label{45}
        \end{eqnarray}
Imaginary parts of these operators are either formally divergent
(beeing the coincidence limits of delta-function type kernels) or
formally zero (as in (\ref{44}) because of vanishing structure functions
components). We shall, however, not ascribe to them particular values
characteristic of gravity theory and keep them of a general form, expecting
that a rigorous operator regularization will exist that can
consistently handle these infinities as well as corresponding quantum
anomalies (see the discussion of this point in \cite{BKr,Barv}).

As shown in \cite{BarvU,Barv}, the DeWitt supermetric on superspace
has as functional Killing vectors the generators of spatial
diffeomorphisms $\nabla^i_a$,
${\cal D}^i\nabla^k_a+{\cal D}^k\nabla^i_a=0,\,
{\cal D}^i=G^{\,ik}{\cal D}_k$, and in view of its ultralocality,
$G^{\,ik}\sim \delta({\bf x}_i,{\bf x}_k)$, the covariant derivative
conserves not only the metric itself, but also the three-point object
$G^{\,ik}_{\perp},\,{\cal D}_m G^{\,ik}_{\perp}=0$. Therefore,
the kinetic terms of operators (\ref{44}) -- (\ref{45}) do not require
additional prescriptions of operator ordering and, moreover, turn out to be
Hermitian with respect to the (unphysical) $L^2$ inner product.
(This does of course not hold for the antihermitean contributions
to the potential term.)

An important result of \cite{GenSem,BarvU,BKr}, which we shall need below,
is a closed construction of the one-loop preexponential factor for a
special two-point solution ${\mbox{\boldmath$K$}}(q,q')$ of the
Wheeler-DeWitt equations represented in the form of the semiclassical ansatz
        \begin{eqnarray}
        {\mbox{\boldmath$K$}}(q,q')={\mbox{\boldmath$P$}}(q,q')
        e^{\,{}^{\textstyle{i m_P^2
        {\mbox{\boldmath$S$}}(q,q')}}}\ .           \label{46}
        \end{eqnarray}
Here, ${\mbox{\boldmath$S$}}(q,q')$ is a particular solution of the
Einstein-Hamilton-Jacobi (EHJ) equations with respect to both arguments --
the classical action calculated at the extremal of equations of motion,
joining the points $q$ and $q'$ in superspace,
        \begin{eqnarray}
        H_\mu(q,\partial {\mbox{\boldmath$S$}}/\partial q)=0.  \label{47}
        \end{eqnarray}

The one-loop ($O(m_P^0)$ part of the $1/m_P^2$-expansion)
 order of the preexponential
factor ${\mbox{\boldmath$P$}}(q,q')$ here satisfies a set of
quasi-continuity equations which follow from the Wheeler-DeWitt
equations at one loop \cite{GenSem,BKr,Barv},
        \begin{eqnarray}
        &&{\cal D}_i(\nabla^i_\mu {\mbox{\boldmath$P$}}^2)=
        U^{\lambda}_{\mu\lambda} {\mbox{\boldmath$P$}}^2,  \label{48}\\
        &&\nabla^i_\mu\equiv\left.
        \frac{\partial H_{\mu}}{\partial p_i}
        \right|_{\textstyle p=\partial
        {\mbox{\boldmath$S$}}/\partial q}\ ,                    \label{49}
        \end{eqnarray}
with the generators $\nabla^i_\mu$ here evaluated at the Hamilton-Jacobi
values of the canonical momenta. The solution of this equation found in
\cite{GenSem,BarvU,BKr} turns out to be a particular generalization of the
Pauli-Van Vleck-Morette formula \cite{Morette} - the determinant calculated
on the subspace of nondegeneracy for the matrix
        \begin{eqnarray}
        {\mbox{\boldmath$S$}}_{ik'}=
        \frac{\partial^2{\mbox{\boldmath$S$}}(q,q')}
        {\partial q^i\, \partial q^{k'}}\ .             \label{50}
        \end{eqnarray}
This matrix has the generators $\nabla^i_\mu$ as zero-eigenvalue eigenvectors
\cite{GenSem,BarvU,BKr}. An invariant algorithm of calculating this
determinant is equivalent to the Faddeev-Popov gauge-fixing procedure
leading to the solution of (\ref{48}). It consists in
introducing a ``gauge-breaking'' term to the matrix (\ref{50}),
        \begin{eqnarray}
        {\mbox{\boldmath$F$}}_{ik'}={\mbox{\boldmath$S$}}_{ik'}+
        \phi^\mu_i c_{\mu\nu} \phi^\nu_{k'}\ ,             \label{51}
        \end{eqnarray}
formed with the aid of gauge-fixing matrix $c_{\mu\nu}$ and two sets of
arbitrary covectors (of ``gauge
conditions'') $\phi^\mu_i$ and $\phi^\nu_{k'}$ at the points $q$ and $q'$
respectively, satisfying invertibility conditions for ``Faddeev-Popov
operators'' at these two points \cite{Fad},
        \begin{eqnarray}
        J^\mu_\nu=\phi^\mu_i\nabla^i_\nu,\,\,
        J\equiv{\rm det}J^\mu_\nu\neq 0,\,\,\,
        J'^{\mu}_{\,\nu}=\phi^\mu_{i'}\nabla^{i'}_\nu,\,\,
        J'\equiv{\rm det}J'^{\mu}_{\,\nu}\neq 0.             \label{52}
        \end{eqnarray}
In terms of these objects the preexponential factor solving the continuity
equations (\ref{48}) is given by the following expression
\cite{GenSem,BarvU,BKr}
        \begin{eqnarray}
        {\mbox{\boldmath$P$}}=
        \left[\frac{{\rm det}{\mbox{\boldmath$F$}}_{ik'}}
        {JJ'\,{\rm det}\,c_{\mu\nu}}\right]^{1/2}           \label{53}
        \end{eqnarray}
which is independent of the introduced arbitrary elements of gauge-fixing
procedure $(\phi^\mu_i,\phi^\nu_{k'}, c_{\mu\nu})$.

\section{Semiclassical ansatz for the two-point solution}
\hspace{\parindent}
We shall now proceed to perform the semiclassical expansion
for solutions to the Wheeler-DeWitt equations.
Since we are interested in giving an interpretation in terms
of Feynman diagrams, we shall not consider wave functionals as
in \cite{Kiefer1}, but two-point solutions (``propagators").
Due to the absence of an external time parameter in the full theory,
such two-point functions play more the role of energy Green functions
than ordinary propagators \cite{Ki91}. However, in the semiclassical
limit, a background time parameter is easily available, with respect
to which Feynman ``propagators" can be formulated.

Let us therefore look for a two-point solution
 of the Wheeler-DeWitt equations in
the form of the ansatz
        \begin{eqnarray}
        \hat{\mbox{\boldmath$K$}}(q_+,q_-)={\mbox{\boldmath$P$}}(q_+,q_-)
        \E^{\,{}^{\textstyle{\I m_P^2
        {\mbox{\boldmath$S$}}(q_+,q_-)}}}
        \hat{\mbox{\boldmath$U$}}(q_+,q_-),           \label{3.1}
        \end{eqnarray}
where we denote (as agreed above) by a hat the operators acting in the Hilbert
space of matter fields\footnote{We note that in \cite{Kiefer1}
$1/D$ corresponds to ${\mbox{\boldmath$P$}}$, while $\chi$
corresponds to $\hat{\mbox{\boldmath$U$}}$.}.
 Here, ${\mbox{\boldmath$S$}}(q_+,q_-)$ is the
principal Hamilton function satisfying the gravitational EHJ (\ref{47})
equations, and ${\mbox{\boldmath$P$}}(q_+,q_-)$ is the preexponential
factor (\ref{53}). Substituting this ansatz into the system of the
Wheeler-DeWitt equations and taking into account the EHJ equations and the
continuity equations for ${\mbox{\boldmath$P$}}(q_+,q_-)$, we get for the
``evolution" operator $\hat{\mbox{\boldmath$U$}}(q_+,q_-)$
the equations
        \begin{eqnarray}
        &&\I\nabla^k_\perp {\cal D}_k\hat{\mbox{\boldmath$U$}}=
        \hat{H}^{\rm mat}_\perp\hat{\mbox{\boldmath$U$}}-
        \frac 1{2m_P^2} {\mbox{\boldmath$P$}}^{-1}
        G^{mn}_\perp {\cal D}_m {\cal D}_n
        ({\mbox{\boldmath$P$}}\,\hat{\mbox{\boldmath$U$}}), \label{3.2}\\
        &&\I\nabla^k_a {\cal D}_k\hat{\mbox{\boldmath$U$}}=
        \hat{H}^{\rm mat}_a\hat{\mbox{\boldmath$U$}},      \label{3.3}
        \end{eqnarray}
where all the derivatives are understood as acting on the argument $q_+$.

Evaluating this operator at the classical extremal $q_+\to q(t)$,
        \begin{eqnarray}
        \hat{\mbox{\boldmath$U$}}(t)=\hat{\mbox{\boldmath$U$}}(q(t),q_-),
        \end{eqnarray}
where $q(t)$ satisfies the canonical equations of motion corresponding to
${\mbox{\boldmath$S$}}(q_+,q_-)$,
        \begin{eqnarray}
        \dot{q^i}=N^\mu \nabla^i_\mu ,               \label{3.3a}
        \end{eqnarray}
one easily obtains the quasi-evolutionary equation
        \begin{eqnarray}
        \I\frac\partial{\partial t}\hat{\mbox{\boldmath$U$}}(t)=
        \hat{H}^{\rm eff}\hat{\mbox{\boldmath$U$}}(t)        \label{3.4}
        \end{eqnarray}
with the {\em effective} matter Hamiltonian
        \begin{eqnarray}
        \hat{H}^{\rm eff}=\hat{H}^{\rm mat}-
        \frac 1{2m_P^2} N^\perp G^{mn}_\perp {\cal D}_m {\cal D}_n
        [\,{\mbox{\boldmath$P$}}\,\hat{\mbox{\boldmath$U$}}\,]
        {\mbox{\boldmath$P$}}^{-1}
        \hat{\mbox{\boldmath$U$}}{\vphantom{U}}^{-1}.     \label{3.5}
        \end{eqnarray}
(Recall that this is an {\em integrated} equation, cf. (2.20).)
The first term on the right-hand side is
 the Hamiltonian of matter fields at the
gravitational background of $(q,N)$-variables $\hat{H}^{\rm mat}$,
        \begin{eqnarray}
        \hat{H}^{\rm mat}=N^\mu\hat{H}^{\rm mat}_\mu\ .       \label{3.6}
        \end{eqnarray}
The second term involves the operator
$\hat{\mbox{\boldmath$U$}}$ itself in a nonlinear way
 and contributes only at order $m_P^{-2}$ of the expansion.
Thus, (\ref{3.4}) is not a true linear Schr\"odinger equation, but
semiclassically it can be solved by iterations starting from the lowest order
approximation
        \begin{eqnarray}
        &&\hat{\mbox{\boldmath$U$}}_0={\rm T}\,{\rm exp}\,
        \left[-i\int_{t_-}^{t_+} \D t
        \hat{H}^{\rm mat}\,\right],                   \label{3.7} \\
        &&\hat{H}^{\rm eff}_0=\hat{H}^{\rm mat}.      \label{3.8}
        \end{eqnarray}
Here, T denotes the Dyson chronological ordering of the usual unitary
evolution operator acting in the Hilbert space of matter fields
$(\hat\varphi,\hat p_\varphi)$. The Hamiltonian
$\hat{H}^{\rm mat}=H^{\rm mat}(\hat\varphi, \hat p_\varphi, q(t),N(t))$
is an operator in the Schr\"odinger picture of these fields
$(\hat\varphi,\hat p_\varphi)$ parametrically depending on gravitational
background variables $(q(t),N(t))$, i.e. evaluated along a {\em particular}
trajectory (``spacetime") in configuration space.

The Dyson T-exponent obviously explains the origin of the standard
Feynman diagrammatic technique in the matter field sector of the theory which
arises in the course of the semiclassical expansion of (\ref{3.7}). We shall
show now
that the gravitational part of this diagrammatic technique involving graviton
loops naturally arises as a result of iterational solution of
(\ref{3.4}) -- (\ref{3.5}) in powers of $1/m_P^2$.

The effective Hamiltonian in the first order approximation of such an
iterational technique can be obtained by substituting (\ref{3.7}) into
(\ref{3.5}) to yield
        \begin{eqnarray}
        &&\hat{H}^{\rm eff}_1(t_+)=\hat{H}^{\rm mat}
        -\frac 1{2m_P^2} {\cal G}^{mn}({\cal D}_m {\cal D}_n
        \hat{\mbox{\boldmath$U$}}_0)
        \hat{\mbox{\boldmath$U$}}_0{\vphantom{U}}^{-1}  \nonumber\\
        &&\qquad\qquad
        -\frac 1{2m_P^2} {\cal G}^{mn}({\cal D}_m {\cal D}_n
        {\mbox{\boldmath$P$}})
        {\mbox{\boldmath$P$}}^{-1}
        -\frac 1{m_P^2} {\cal G}^{mn}
        ({\cal D}_m{\mbox{\boldmath$P$}}){\mbox{\boldmath$P$}}^{-1}\,
        ({\cal D}_n\hat{\mbox{\boldmath$U$}}_0)
        \hat{\mbox{\boldmath$U$}}{}^{-1}_0.    \label{3.9}
        \end{eqnarray}
Here we have used the new notation
        \begin{eqnarray}
        {\cal G}^{mn}=N^\perp G^{mn}_\perp      \label{3.9a}
        \end{eqnarray}
for another metric on the configuration space
 (compare with (\ref{38})) which uses
the actual value of the lapse function corresponding to the classical extremal
(\ref{3.3a}). This lapse function
 generally differs from unity. We also decomposed the
first-order corrections in the effective Hamiltonian into three terms
corresponding to the contribution of quantum matter (generated by
$\hat{\mbox{\boldmath$U$}}_0$), purely quantum gravitational contribution
generated by ${\mbox{\boldmath$P$}}$, and their cross term\footnote{The
second correction term in (\ref{3.9}) was ``absorbed" in \cite{Kiefer1}
by the introduction of some wave functional $\sigma$ obeying a
second-order WKB equation.}.

Further evaluation of these terms demands the knowledge of derivatives
acting on the configuration space argument $q_+$ of $\hat{\mbox{\boldmath$U$}}_0$ and
${\mbox{\boldmath$P$}}$. Obtaining
these derivatives leads to the necessity of considering the special
boundary value problem for classical equations of motion, the graviton
propagator and vertices -- elements of the gravitational Feynman diagrammatic
technique. The rest of the paper will be mainly devoted to present
the corresponding derivations.

We begin by introducing the collective notation for the full set of
{\em Lagrangian} gravitational variables, which includes both the spatial
metric as well as lapse and shift functions,
        \begin{eqnarray}
        &&g^a\equiv(q^i(t),N^\mu(t))\ .     \label{3.10}
        \end{eqnarray}
Actually this collection of ten fields $g^a\sim g_{\alpha\beta}({\bf x},t)$
comprises the whole set of spacetime metric coefficients taken in a special
parametrization adjusted to the (3+1)-splitting. In what follows we shall need
also spacetime condensed DeWitt notations in which the index $a$ includes not
only the spatial coordinates ${\bf x}$ but also the time variable $t$, and
contraction of these indices will imply also the integration over $t$.
 In spacetime
condensed notations the gravitational action has the form
        \begin{eqnarray}
        S[\,g\,]=\int_{t_-}^{t_+}\D t\,L(q(t),\dot{q}(t),N^\mu(t))\ .
         \label{3.12}
        \end{eqnarray}
The Lagrangian does not involve the time derivatives of lapse
and shift functions $N^\mu(t)$ which play the role of
 Lagrange multipliers for
the first-class constraints. This Lagrangian is related to
the integrand of the canonical action (\ref{34})
        \begin{eqnarray}
        L(q,\dot{q},N^\mu)=\left.\left( p_i\dot{q}^i-
        N^{\mu}H_{\mu}(q,p)\right)\right|_{p=p^0(q,\dot{q},N)}  \label{3.12a}
        \end{eqnarray}
by the substitution of the expression for the canonical momentum
$p_i^0(q,\dot{q},N)$ in terms of the velocities $\dot{q}$ -- the
solution of the canonical equation of motion
        \begin{eqnarray}
        \dot{q}{}^i=N^\mu\frac{\partial H_\mu}{\partial p_i}\ .  \label{3.12b}
        \end{eqnarray}

In the notations of the above type the evolution operator
(\ref{3.7}) as a function of $q_+$ can be regarded
as a functional of the classical extremal $g^a=(q^i(t),N^\mu(t))$ joining the points
$q_+$ and $q_-$ in superspace at the respective
 moments of time $t_+$ and $t_-$,
        \begin{eqnarray}
        \hat{\mbox{\boldmath$U$}}_0(q_+,q_-)=
        \hat{\mbox{\boldmath$U$}}_0[\,g(t|q_+,q_-)\,].    \label{3.11}
        \end{eqnarray}
Here, $g(t|q_+,q_-)$ is a classical extremal of vacuum Einstein equations
parametrically depending on end points $q_\pm$ at $t_\pm$. In terms of the
action (\ref{3.12}) the boundary value problem for this extremal
can be written as
        \begin{eqnarray}
        &&\frac{\delta S[g]}{\delta g^a(t)}=0,    \label{3.13}\\
        &&q(t_\pm)=q_\pm\ .                         \label{3.14}
        \end{eqnarray}

Thus, the derivative of $\hat{\mbox{\boldmath$U$}}_0$ with
respect to $q_+$ can be written as
        \begin{eqnarray}
        \frac{\partial\hat{\mbox{\boldmath$U$}}_0}{\partial q_+^m}=
        \int_{t_-}^{t_+} \D t\, \frac{\partial g^a(t)}{\partial q_+^m}
        \frac{\delta\hat{\mbox{\boldmath$U$}}_0[\,g\,]}{\delta g^a(t)}
        \equiv \frac{\partial g^a}{\partial q_+^m}
        \frac{\delta\hat{\mbox{\boldmath$U$}}_0}{\delta g^a}. \label{3.15}
        \end{eqnarray}
Note that in the last equality we used the contraction rule for spacetime
condensed indices, implying the time integration together with the spatial
integration. In what follows we shall not introduce special labels to
distinguish between condensed spatial and spacetime notations. As a rule,
when the time argument is explicitly written, we shall imply that the
corresponding condensed index or indices involve only spatial coordinates and
their contraction does not involve implicit time integrals. Another
distinction between these two types of condensed notations concerns functional
derivatives. We shall always reserve functional variational notations
$\delta/\delta g^a\equiv\delta/\delta g^a(t)$ for variational derivatives
with respect to functions of time, while the variational derivatives with
respect to functions of spatial coordinates will be denoted by partial
derivatives $\partial/\partial q^i\equiv\delta/\delta g_{ab}({\bf x})$.

By the same method we shall calculate $q_+$-derivatives of the gravitational
prefactor ${\mbox{\boldmath$P$}}$ and the higher-order derivatives of all
the quantities in
question. An important ingredient of all these calculations is the quantity
        \begin{eqnarray}
        \frac{\partial g^a(t|q_+,q_-)}{\partial q_+^m}\ .      \label{3.16}
        \end{eqnarray}
It will be obtained from the following linear boundary value problem.

\section{Linear boundary value problem}
\hspace{\parindent}
To find (\ref{3.16}) one has to consider the boundary value problem
(\ref{3.13}) -- (\ref{3.14}) and make an arbitrary infinitesimal variation of
the final point of the
extremal $\delta q^i_+$. This variation will induce the variation of the
extremal $\delta q_+\neq 0 \rightarrow \delta g^a=(\delta q^i(t),
\delta N^\mu(t))$ satisfying the linear boundary value problem
        \begin{eqnarray}
        &&S_{ab}\delta g^b=0,\,\,\,
        S_{ab}\equiv\frac{\delta^2 S[g]}
        {\delta g^a(t)\delta g^b(t')},                 \label{4.1}\\
        &&\delta q(t_+)=\delta q_+, \quad \delta q(t_-)=0 \ .   \label{4.2}
        \end{eqnarray}
This follows from (4.18) after taking account that the varied trajectory
must also be an extremum.
The solution of this problem obviously gives the needed quantity (\ref{3.16}).
However, the solution is not unique because of the
local gauge invariance of the gravitational action (\ref{3.12}).

This gauge invariance with respect to spacetime diffeomorphisms, discussed
above in the canonical formalism, in the Lagrangian formalism
takes the form of the invariance with respect to infinitesimal
transformations with local parameters $f^\mu=f^\mu(t)$ and generators
$R^a_\mu=(R^i_\mu,R^\alpha_\mu)$. In view of the relation (\ref{3.12a})
between the canonical and Lagrangian formalisms these gauge transformations
can be obtained from the invariance transformations (\ref{40}) -- (\ref{41})
\cite{BFV}
        \begin{eqnarray}
        &&g^a\rightarrow g^a+R^a_\mu f^\mu,    \label{4.3}\\
        &&R^i_\mu=\delta(t_i-t_\mu)\left.\frac{\partial H_\mu}{\partial p_i}
        \,\right|_{p=p^0(q,\dot{q},N)},\,\,\,
        \left.R^i_\mu\vphantom{\frac\partial\partial}
        \,\right|_{p=\partial S/\partial q}=
        \nabla^i_\mu\,\delta(t_i-t_\mu),                    \label{4.4}\\
        &&R^\alpha_\mu=\left(\delta^\alpha_\mu\frac d{dt_\alpha}
        -U^\alpha_{\lambda\mu}
        N^\lambda\right)\delta(t_\alpha-t_\mu).             \label{4.5}
        \end{eqnarray}
Note that in the definition of gauge generators we used condensed notations
in which they have the above form of delta-function type kernels with
subscripts of time coordinates indicating to which spacetime condensed index
they belong. Note also that gauge transformations of superspace coordinates
are ultralocal in time in contrast with the Lagrange multipliers whose
transformations include the time derivative of the gauge parameter,
see (3.11). The
invariance of action in terms of these generators takes the form of the
following identities
        \begin{eqnarray}
        R^a_\mu\frac{\delta S[g]}{\delta g^a}=0.          \label{4.6}
        \end{eqnarray}
Their functional differentiation shows that on-shell, that is on the solution
of classical equations of motion $S_a=0$, the Hessian of the action $S_{ab}$
is not an invertible operator because it has zero-eigenvalue eigenvectors --
the gauge generators:
        \begin{eqnarray}
        R^a_\mu S_{ab}=-S_a\frac{\delta R^a_\mu}{\delta g^b}=0. \label{4.7}
        \end{eqnarray}
This is of course a generic feature of
constrained systems.
The degeneracy implies that there is no unique solution to the linear
problem (\ref{4.1}) for $\delta g$, which is always defined up to gauge
transformations
of the above type. The standard method of singling out a particular solution
from their gauge equivalence class uses the gauge-fixing procedure
\cite{DW,Fad}. This procedure in application to the full nonlinear equations
of motion (\ref{3.13}) -- (\ref{3.14}) consists in imposing gauge conditions
        \begin{eqnarray}
        \chi^\mu(g,\dot{g})=0,                    \label{4.8}
        \end{eqnarray}
which should have a nondegenerate Faddeev-Popov operator (or operator of
ghost fields) -- the two-point kernel of their gauge transformation
        \begin{eqnarray}
        &&Q^\mu_\nu\equiv\frac{\delta\chi^\mu}
        {\delta g^a}R^a_\nu,                        \label{4.9}\\
        &&{\rm det}\,Q^\mu_\nu\neq 0.               \label{4.9a}
        \end{eqnarray}

For the linearized problem (\ref{4.1}) the gauge fixing can obviously
be reduced to imposing the linearized gauge condition on field
disturbances $\delta g$,
        \begin{eqnarray}
        \chi^\mu_a \delta g^a=0,\,\,\,\chi^\mu_a
        \equiv\frac{\delta\chi^\mu}{\delta g^a}.             \label{4.10}
        \end{eqnarray}
In view of this condition, the linear equation (\ref{4.1}) on $\delta g$
can be replaced by another equation
        \begin{eqnarray}
        &&F_{ab}\delta g^b=0,                  \label{4.11}\\
        &&F_{ab}\equiv S_{ab}-\chi^\mu_a c_{\mu\nu} \chi^\nu_b, \label{4.12}
        \end{eqnarray}
where the operator $F_{ab}$ is supplied by a gauge-breaking term with some
invertible gauge-fixing matrix $c_{\mu\nu}$,
cf. (3.21). In contrast with $S_{ab}$, this
operator is nondegenerate because now
        \begin{eqnarray}
        R^a_\mu F_{ab}=-Q^\alpha_\mu
        c_{\alpha\beta} \chi^\beta_b,               \label{4.13}
        \end{eqnarray}
and the right-hand side represents here in view of (\ref{4.9a}) a
functional matrix with the rank coinciding with the dimensionality of
the gauge group.

Now we have to consider briefly the properties of the obtained operator
$F_{ab}$. Note that the original operator $S_{ab}$ was of second order in
time derivatives only in the sector of $q^i$-variables,
$S_{ik}=\delta^2 S/\delta q^i(t)\delta q^k(t')$, because only these variables
enter the gravitational Lagrangian with their velocities $\dot{q}$.
Correspondingly, the mixed sector $S_{i\mu}=\delta^2 S/\delta q^i\delta N^\mu$
is a first-order differential operator in time, while
$S_{\mu\nu}=\delta^2 S/\delta N^\mu\delta N^\nu$ is ultralocal in time, that
is proportional to an undifferentiated delta-function,
$S_{\mu\nu}\sim\delta(t_\mu-t_\nu)$. The addition of a gauge-breaking term in
$F_{ab}$ can drastically change this structure,
 provided the chosen gauge conditions
(\ref{4.8}) are so-called {\em relativistic} ones. Relativistic gauges
contain time derivatives of lapse and shift functions so that,
        \begin{eqnarray}
        \mbox{det} \frac{\partial\chi^\mu}
        {\partial\dot{N}^\nu}\neq 0\ ,                \label{4.14}
        \end{eqnarray}
and the corresponding Faddeev-Popov operator is of second order in time
derivatives. (This is a simple consequence of the fact
 that the gauge generator
(\ref{4.5}) in the sector of Lagrange multipliers is itself a first-order
differential operator.) It reads
        \begin{eqnarray}
        Q^\mu_\nu=\frac{\partial\chi^\mu}{\partial\dot{N}^\nu}
        \frac{d^2}{dt^2_\mu}\delta(t_\mu-t_\nu)+\ldots \, .     \label{4.15}
        \end{eqnarray}
This choice of gauge conditions guarantees that the functional matrix
$\chi^\mu_a$ of linearized gauge conditions is also a first-order differential
operator with the delta-function type kernel
        \begin{eqnarray}
        \chi^\mu_a=\stackrel{\rightarrow}{\chi}
        {\!}^\mu_a(d/dt_\mu)
        \delta(t_\mu-t_a).                    \label{4.16}
        \end{eqnarray}
In what follows we shall often denote the differential operators and the
direction (right or left) in which they act by arrows.

Thus, with the choice of relativistic gauge conditions the operator $F_{ab}$
is of second order in time derivatives acting in all sectors of test
fields $\varphi^a=(\varphi^i,\varphi^\mu)$. For any two such test fields
$\varphi_1$ and $\varphi_2$ one can write down the Wronskian relations for
the original Hessian operator
$S_{ab}=\stackrel{\rightarrow}{S}_{ab}(d/dt)\delta(t-t')$
        \begin{eqnarray}
        &&\varphi^a_1(t)\stackrel{\rightarrow}{S}_{ab}\varphi^b_2(t)
        -\varphi^a_1(t)\stackrel{\leftarrow}{S}_{ab}\varphi^b_2(t)
        =\nonumber\\
        &&\qquad\qquad\qquad\qquad\qquad\frac d{dt}\left[\varphi^a_1(t)
        \stackrel{\rightarrow}{W}_{ab}\varphi^b_2(t)-
        \varphi^a_1(t)\stackrel{\leftarrow}{W}_{ab}
        \varphi^b_2(t)\right]                           \label{4.17}
        \end{eqnarray}
and for the gauge-fixed operator
$F_{ab}=\stackrel{\rightarrow}{F}_{ab}(d/dt)\delta(t-t')$
        \begin{eqnarray}
        &&\varphi^a_1(t)\stackrel{\rightarrow}{F}_{ab}\varphi^b_2(t)
        -\varphi^a_1(t)\stackrel{\leftarrow}{F}_{ab}\varphi^b_2(t)
        =\nonumber\\
        &&\qquad\qquad\qquad\qquad\qquad\frac d{dt}\left[\varphi^a_1(t)
        \stackrel{\rightarrow}{W}{\!}^F_{ab}\varphi^b_2(t)-
        \varphi^a_1(t)\stackrel{\leftarrow}{W}{\!}^F_{ab}
        \varphi^b_2(t)\right].                             \label{4.18}
        \end{eqnarray}
In the {\em Wronskian} operator $\stackrel{\rightarrow}{W}_{ab}(d/dt)$ only
the components
$\stackrel{\rightarrow}{W}_{ib}$ are nonvanishing.
They can be obtained from the gravitational
Lagrangian by varying the canonical momentum conjugate to $q^i$
        \begin{eqnarray}
        \stackrel{\rightarrow}{W}_{ib}(d/dt)\delta g^b(t)=
        -\delta\,\frac{\partial L(q,\dot{q},N)}
        {\partial\dot{q}^i}.                           \label{4.19}
        \end{eqnarray}
For the operator $F_{ab}$ all the components of its Wronskian operator
are nonvanishing and related to $\stackrel{\rightarrow}{W}_{ab}$ by
a simple equation,
        \begin{eqnarray}
        \stackrel{\rightarrow}{W}{\!}^F_{ab}(d/dt)=
        \stackrel{\rightarrow}{W}_{ab}(d/dt)+
        \frac{\partial\chi^\mu}{\partial\dot{g}^a}c_{\mu\nu}
        \stackrel{\rightarrow}{\chi}{\!}^\nu_b(d/dt).     \label{4.20}
        \end{eqnarray}

Now we can construct the solution of the boundary value problem (\ref{4.2}),
(\ref{4.10}) and (\ref{4.11})
for a small disturbance $\delta g^a$ of the classical extremal induced by a
variation of its end point $\delta q^i_+$. First, let us introduce a
special Green function $D^{ab}(t,t')$ of the operator $F_{ab}$
        \begin{eqnarray}
        &&F_{ab}D^{bc}=\delta^c_a,                       \label{4.21}\\
        &&D^{ic}(t_\pm,t)=0,                             \label{4.22}\\
        &&\stackrel{\rightarrow}{\chi}{\!}^\mu_b(d/dt_\pm)
        D^{bc}(t_\pm,t)=0.                               \label{4.23}
        \end{eqnarray}
This Green function is uniquely defined by the Dirichlet boundary conditions
on its $q^i$-components (\ref{4.22}) and by special Robin-type boundary
conditions (\ref{4.23}). It plays the role of a graviton propagator analogous
to the causal Feynman
Green function, but with special boundary conditions appropriate to the
definition of the two-point solution of the Wheeler-DeWitt equation (\ref{46}).
It was first introduced in \cite{barvin} where it was shown that this is a
propagator of the semiclassical loop expansion of this solution.
 Its boundary
conditions have a number of remarkable properties
 including the BRST-invariance
and selfadjointness of the graviton operator \cite{aveskam}. Here we add one
more property which is proved in Appendix~A:
 this Green function gives a solution
to our linear problem for $\delta g^a$ induced by $\delta q^i_+$,
        \begin{eqnarray}
        \delta g^a(t)=\delta q^i_+\stackrel{\rightarrow}
        {W}_{ib}\!(d/dt_+)D^{ba}(t_+,t).                     \label{4.24}
        \end{eqnarray}
{}From this follows immediately the desired answer
 for the derivative (\ref{3.16}),
        \begin{eqnarray}
        \frac{\partial g^a(t)}{\partial q^i_+}=
        \stackrel{\rightarrow}{W}_{ib}\!D^{ba}(t_+,t).   \label{4.25}
        \end{eqnarray}
The perturbation scheme can now be employed.

\section{Heisenberg operators of matter fields}
\hspace{\parindent}
Now we can continue the calculation of the effective Hamiltonian (\ref{3.9})
in the modified Schr\"odinger equation (\ref{3.4}). Finding the derivative
(\ref{3.15}) requires
the know\-ledge of the functional derivatives of the unitary evolution operator
$\hat{\mbox{\boldmath$U$}}_0$ which is rather straightforward. In view of the
chronologically ordered nature of the latter one has the expression (time
arguments of evolution operators denote the time intervals in
which they are acting)
        \begin{eqnarray}
        \frac{\delta\hat{\mbox{\boldmath$U$}}_0}{\delta g^a(t)}=
        -\I\,\hat{\mbox{\boldmath$U$}}_0(t_+,t)
        \frac{\partial\hat{H}^{\rm mat}}{\partial g^a(t)}
        \hat{\mbox{\boldmath$U$}}_0(t,t_-).                    \label{5.1}
        \end{eqnarray}
This can be rewritten in terms of the Heisenberg operator of the matter
Hamiltonian, $\hat{H}^{\rm mat}_h(t)$, as
        \begin{eqnarray}
        \frac{\delta\hat{\mbox{\boldmath$U$}}_0}{\delta g^a(t)}=
        -\I\,\frac{\partial\hat{H}^{\rm mat}_h(t)}{\partial g^a(t)}
        \hat{\mbox{\boldmath$U$}}_0.                         \label{5.2}
        \end{eqnarray}
Here the Heisenberg operator $\hat{H}^{\rm mat}_h(t)$ differs from the
Hamiltonian $\hat{H}^{\rm mat}$ in the Schr\"o\-ding\-er picture
        \begin{eqnarray}
        \hat{H}^{\rm mat}_h(t)=\hat{\mbox{\boldmath$U$}}_0(t_+,t)
        \hat{H}^{\rm mat}\hat{\mbox{\boldmath$U$}}{}^{-1}_0(t_+,t)=
        \hat{H}^{\rm mat}
        (\hat\varphi(t), \hat p_\varphi(t), g(t))             \label{5.3}
        \end{eqnarray}
by replacing the Schr\"odinger matter-field operators,
$(\hat\varphi, \hat p_\varphi)$, with the Heisenberg ones,
$(\hat\varphi(t), \hat p_\varphi(t))$. The Heisenberg operators satisfy
operator equations of motion. Therefore, the derivative of the
Hamiltonian in (\ref{5.2}) with respect to the gravitational variable
can be replaced by the functional derivative of the matter field action
$\hat{S}{}^{\rm mat}$ (also taken in the Heisenberg picture)\footnote
{Indeed, this follows from differentiating the relation $L^{\rm mat}=
p_\varphi\dot{\varphi}-H^{\rm mat}$ which gives extra terms proportional
to the equations of motion for $\varphi$. Up to some operator-ordering
ambiguity (usually absorbed by renormalization in renormalizable matter
field theories), these terms vanish for Heisenberg operators satisfying
these equations.
}
        \begin{eqnarray}
        \frac{\partial\hat{H}^{\rm mat}_h(t)}{\partial g^a(t)}=
        -\frac{\delta\hat{S}^{\rm mat}}{\delta g^a(t)}\equiv
        -\hat{T}_a(t).                                       \label{5.4}
        \end{eqnarray}
But the right-hand side of this relation is just the matter stress
tensor in the Heisenberg picture of matter fields decomposed in the normal
basis of (3+1)-foliation. Going back from condensed notations to the usual
ones, it can easily be checked that
        \begin{eqnarray}
        \hat{T}_a=\frac{\delta\hat{S}^{\rm mat}}{\delta g_{\alpha\beta}(x)}
        \frac{\partial g_{\alpha\beta}}{\partial g^a}=
        \left(\,\frac 12 T^{ab}(x),\,
        \frac 1{N^\perp} T_{\perp a}(x),\,
        -\frac 1{N^\perp} T_{\perp\perp}(x)\,\right),       \label{5.5}
        \end{eqnarray}
where
        \begin{eqnarray}
        T^{\alpha\beta}(x)=
        2\frac{\delta S^{\rm mat}}{\delta g_{\alpha\beta}(x)}  \label{5.6}
        \end{eqnarray}
is a conventional matter stress tensor (density)\footnote
{
The condensed notation (\ref{5.5}) should not be
mixed up with the notation (\ref{45}) for the total momentum constraint in the
canonical formalism. Since the latter will not be used below we take the
liberty to reserve this notation for a spacetime covariant quantity --
the stress tensor of matter fields.
}.

Thus we finally have
        \begin{eqnarray}
        \frac{\delta\hat{\mbox{\boldmath$U$}}_0}{\delta g^a(t)}=
        \I\,\hat{T}_a(t)\hat{\mbox{\boldmath$U$}}_0.         \label{5.7}
        \end{eqnarray}
The substitution of this result into the $q_+$-derivative of the evolution
operator (\ref{3.15}) gives rise to the quantity
        \begin{eqnarray}
        \hat{t}{}^a(t)=-\frac 1{m_P^2}\int_{t_-}^{t_+} \D t'\,
        D^{ab}(t,t')\hat{T}_b(t')\equiv
        -\frac 1{m_P^2}\,D^{ab}\hat{T}_b\ ,            \label{5.10}
        \end{eqnarray}
where we have again introduced a condensed notation in the last step.
This quantity is a solution of the linear inhomogeneous equation
 (see Appendix~A)
        \begin{eqnarray}
        m_P^2\,\stackrel{\rightarrow}{S}_{ab}\hat{t}{}^b(t)+
        \hat{T}_a(t)=0,                                \label{5.11}
        \end{eqnarray}
which are the linearized {\em nonvacuum} Einstein equations with operator
matter source. Therefore, $\hat{t}{}^a(t)$ can be regarded as a gravitational
potential generated by the back reaction of quantum matter on the
gravitational background. It satisfies the same boundary conditions at
$t_\pm$ as the Green function $D^{ab}$ and the same linearized gauge
conditions
        \begin{eqnarray}
        \chi^\mu_a\hat{t}{}^a=0.                       \label{5.12}
        \end{eqnarray}
As shown in Appendix A, these properties follow from the covariant
conservation law for the Heisenberg operator of matter stress tensor
(\ref{5.5}). This
conservation is in turn a consequence of the diffeomorphism invariance
of the matter action and Heisenberg equations of motion for matter fields
        \begin{eqnarray}
        R^a_\mu\hat{T}_a=-R^\varphi_\mu
        \frac{\delta\hat{S}^{\rm mat}}{\delta\varphi}=0   \label{5.13}
        \end{eqnarray}
(here $R^\varphi_\mu$ are the generators of local gauge transformations of
matter fields -- their local diffeomorphisms).

In terms of the constructed gravitational potential, the $q_+$-derivative of
the evolution operator (\ref{3.15}) takes the form
        \begin{eqnarray}
        \frac{\partial\hat{\mbox{\boldmath$U$}}_0}
        {\partial q_+^m}={\cal D}_m\hat{\mbox{\boldmath$U$}}_0=
        -\I\,m_P^2\,\stackrel{\rightarrow}{W}_{ma}\!\hat{t}{}^a(t_+)
        \hat{\mbox{\boldmath$U$}}_0.                    \label{5.14}
        \end{eqnarray}

The second-order (covariant) $q_+$-derivative of
 $\hat{\mbox{\boldmath$U$}}_0$
which is contained in the effective Hamiltonian (\ref{3.9})
-- the first correction term proportional to $m_P^{-2}$ -- follows
from differentiating (4.20) and using (6.7). It equals
        \begin{eqnarray}
        {\cal D}_m {\cal D}_n\hat{\mbox{\boldmath$U$}}_0
        =\I\left({\cal D}_m\frac{\partial g^a}{\partial q^n_+}\right)
        \hat{T}_a\hat{\mbox{\boldmath$U$}}_0
        +\frac{\partial g^a}{\partial q^m_+}\,
        \frac{\partial g^b}{\partial q^n_+}\,
        \frac{\delta^2\hat{\mbox{\boldmath$U$}}_0}
        {\delta g^a\delta g^b}.                            \label{5.15}
        \end{eqnarray}
The second-order variational derivative of the evolution operator here
can be obtained by repeated functional differentiation of (\ref{5.7})
to give the expression
        \begin{eqnarray}
        &&\frac{\delta^2\hat{\mbox{\boldmath$U$}}_0}{\delta g^a\delta g^b}=
        \I\frac{\delta^2\hat{S}{}^{\rm mat}}{\delta g^a\delta g^b}
        \hat{\mbox{\boldmath$U$}}_0-
        {\rm T}\left(\hat{T}_a\hat{T}{}_b\right)
        \hat{\mbox{\boldmath$U$}}_0,                     \label{5.8}\\
        &&{\rm T}\left(\hat{T}_a\hat{T}_b\right)
        =\theta(t_a-t_b)\hat{T}_a\hat{T}_b
        +\theta(t_b-t_a)\hat{T}_b\hat{T}_a,         \label{5.9}
        \end{eqnarray}
where T denotes the operator chronological ordering.

Similarly, the first term in (\ref{5.15}) follows from repeated
differentiation of (\ref{4.25}) and expressing the answer in terms of
functional variations of the gravitational background. In view of the
variational equation for the Green function $D^{ab}$
        \begin{eqnarray}
        \delta D^{ab}=-D^{ac}\delta F_{cd}D^{db}=
        -D^{ac}S_{cde}D^{db}\delta g^e,
        \end{eqnarray}
where\footnote
{
We assume for simplicity that the matrix of gauge conditions $\chi^\mu_a$
can be chosen background field independent, whence
$\delta F_{cd}=S_{cde}\delta g^e$. Otherwise extra vertices involving
the variational derivatives $\delta\chi^\mu_a/\delta g^b$ will appear
and generate an extra set of diagrams with ghost propagators of the
Faddeev-Popov operator (\ref{4.9}).
}
        \begin{eqnarray}
        S_{cde}\equiv\frac{\delta^3 S[g]}
        {\delta g^c\delta g^d\delta g^e}
        \end{eqnarray}
is a three-point gravitational vertex, one has
        \begin{eqnarray}
        \left({\cal D}_m\frac{\partial g^a}{\partial q^n_+}\right)
        \hat{T}_a=-m_P^2({\cal D}_m\!
        \stackrel{\rightarrow}{W}_{na})\,\hat{t}{}^a
        +m_P^2(\stackrel{\rightarrow}{W}_{ma}\!D^{ac})
        (\stackrel{\rightarrow}{W}_{nb}\!D^{bd})
        S_{cde}\,\hat{t}{}^e.                          \label{5.16}
        \end{eqnarray}

Combining (\ref{5.15}) with (\ref{5.8}) and (\ref{5.16}) and contracting
these expressions with ${\cal G}^{mn}$ one can get the contribution of matter
fields to the effective Hamiltonian (\ref{3.9}).
It contains, however, the object
 $({\cal G}^{mn}{\cal D}_m\!\stackrel{\rightarrow}{W}_{na})\,\hat{t}{}^a$
which requires further simplification.
 The Wronskian operator here is a local
quantity taken at the moment of time $t_+$, but its $q_+$-derivative
calculated as above in terms of the background field variation,
$\partial/\partial q_+=(\partial g^a/\partial q_+)\delta/\delta g^a$,
again produces a nonlocality -- the Green function $D^{ab}(t,t')$ and
its derivatives with respect to both arguments $t$ and $t'$ taken at the
coincident points $t=t'=t_+$. This can be condensely written down
in the form of a special local three-point vertex $w_{abc}(t_+)$
        \begin{eqnarray}
        ({\cal G}^{mn}{\cal D}_m\!\stackrel{\rightarrow}{W}_{na})
        \,\hat{t}{}^a\,&=&\,
        D^{ab}w_{abc}(t_+)\,\hat{t}{}^c  \nonumber\\
        \,&\equiv&\,
        \left.D^{ab}(t,t')\stackrel{\leftrightarrow}
        {w}_{abc
}(d/dt,d/dt',d/dt'')\,
        \hat{t}{}^c(t'')\right|_{t=t'=t''=t_+}.        \label{5.17}
        \end{eqnarray}
The form of $w_{abc}(t_+)$ is rather complicated and can be found in 
Appendix B.

With these notations the final form of the matter field contribution to
$\hat{H}{}^{\rm eff}_1$ -- the first correction term -- reads
        \begin{eqnarray}
        &&-\frac 1{2m_P^2} {\cal G}^{mn}({\cal D}_m {\cal D}_n
        \hat{\mbox{\boldmath$U$}}_0)
        \hat{\mbox{\boldmath$U$}}_0{}^{-1}=
        \frac 12 m_P^2\,{\cal G}^{mn}\,{\rm T}
        \left(\stackrel{\rightarrow}{W}_{ma}\!\hat{t}{}^a\,
        \stackrel{\rightarrow}{W}_{nb}\!\hat{t}{}^b\right)
        +\frac{\I}{2} D^{ab}w_{abc}(t_+)\,\hat{t}{}^c\nonumber\\
        &&\qquad\qquad\qquad-\frac{\I}{2} {\cal G}^{mn}\,
        (\stackrel{\rightarrow}{W}_{ma}\!D^{ac})\,
        (\stackrel{\rightarrow}{W}_{nb}\!D^{bd})(\,S_{cde}\,
        \hat{t}{}^e
        +\frac 1{m_P^2}\hat{S}{}^{\rm mat}_{cd}\,)\ .     \label{5.18}
        \end{eqnarray}

The resulting three terms can be given a Feynman diagrammatic representation
with different structure. Note that because of (6.8) all terms
are of the same order $m_P^{-2}$.
 The first term begins with the tree-level
structure quadratic in gravitational potential operators $\hat{t}{}^a$.
Note that despite the fact that these operators are taken at one moment of
time $t_+$, their chronological product is nontrivial because it should
read as
        \begin{eqnarray}
        {\rm T}\left(\hat{t}{}^a\,\hat{t}{}^b\right)= \frac 1{m_P^4}\,
        D^{ac}D^{bd}\,{\rm T}\left(\hat{T}_c\,\hat{T}_d\right)
        \end{eqnarray}
and, thus, includes all higher order chronological couplings between
composite operators of matter stress tensors. The second and the third terms
in (\ref{5.18}) are essentially quantum, because their semiclassical
expansions start with the one-loop diagrams consisting of one and two
``graviton propagators" $D^{ab}$, respectively. The quasi-local vertices of
these diagrams are built from the Wronskian
operators, gravitational
three-vertices $w_{abc}(t_+)$ and $S_{cde}$ and second-order variation of
matter action with respect to gravitational variables
$\hat{S}{}^{\rm mat}_{cd}$. The corresponding diagrams are shown in Fig.~1.

\section{Graviton loop effects}
\hspace{\parindent}
The quantum gravitational contribution to the effective Hamiltonian
(\ref{3.9}) is generated by the gravitational preexponential factor
${\mbox{\boldmath$P$}}(q_+,q_-)$. In the one-loop approximation it is
known as the solution (\ref{53}) of the continuity equation (\ref{48})
which itself follows from
the Wheeler-DeWitt equations. On the other hand, within the path integral
representation of this solution one can write down the spacetime covariant
representation of this quantity as a one-loop (gaussian) approximation for
the path integral. Then, in contrast to the spatial functional determinants
of (\ref{53}), it will be given by {\em spacetime} functional determinants of
differential wave operators of gravitational and ghost fields
(in a certain gauge-fixing procedure). This
problem was solved in \cite{barvin}, where it was shown that the prefactor
is given by the one-loop effective action of the theory calculated on the
classical extremal (\ref{3.13}) -- (\ref{3.14}) joining the superspace points
$q_\pm$ at some moments $t_\pm$,
        \begin{eqnarray}
        &&{\mbox{\boldmath$P$}}(q_+,q_-)=
        \E^{\,{}^{\textstyle{\I{\mbox{\boldmath $\Gamma$}}}
        [\,g(t|q_+,q_-)\,]}},                           \label{6.1}\\
        &&{\mbox{\boldmath $\Gamma$}}=\frac{\I}{2}\, {\rm Tr}\ln F_{ab}-
        \I\, {\rm Tr}\ln Q^\mu_\nu\ .                         \label{6.2}
        \end{eqnarray}
A remarkable property of this effective action
${\mbox{\boldmath $\Gamma$}}[g]$ is that its gauge ($F_{ab}$) and ghost
($Q^\mu_\nu$) inverse propagators coincide with the operators (\ref{4.12})
and (\ref{4.9}) introduced above. Moreover, their functional determinants are
calculated
with the same boundary conditions as those of the Green function
$D^{ab}$ (\ref{4.22}) -- (\ref{4.23})
 and Dirichlet boundary conditions of the
ghost Green function $Q^{-1\mu}_{\,\nu}$ \cite{barvin}. Therefore, to
calculate the graviton contribution to $\hat{H}{}^{\rm eff}_1$ one can repeat
the steps of the previous section and arrive at the equations similar to
(\ref{5.14}),
        \begin{eqnarray}
        {\cal D}_m{\mbox{\boldmath$P$}}=
        -\I\,m_P^2\,\stackrel{\rightarrow}{W}_{ma}\gamma^a(t_+)
        {\mbox{\boldmath$P$}}\ ,                            \label{6.2a}
        \end{eqnarray}
and (\ref{5.18}),
        \begin{eqnarray}
        &&-\frac 1{2m_P^2} {\cal G}^{mn}({\cal D}_m {\cal D}_n
        {\mbox{\boldmath$P$}}){\mbox{\boldmath$P$}}^{-1}=
        \frac 12 m_P^2\,{\cal G}^{mn}\,
        (\stackrel{\rightarrow}{W}_{ma}\!\gamma^a)
        (\stackrel{\rightarrow}{W}_{nb}\!\gamma^b)
        +\frac{\I}{2} D^{ab}w_{abc}(t_+)\,\gamma^c     \nonumber\\
        &&\qquad\qquad\qquad-\frac{\I}{2} {\cal G}^{mn}\,
        (\stackrel{\rightarrow}{W}_{ma}\!D^{ac})\,
        (\stackrel{\rightarrow}{W}_{nb}\!D^{bd})(\,S_{cde}\,
        \gamma^e
        +\frac 1{m_P^2}{\mbox{\boldmath $\Gamma$}}_{cd}\,).  \label{6.3}
        \end{eqnarray}
Comparing this with (6.20), the main modification is the replacement
of the gravitational potential
$\hat{t}{}^a$ and the matter vertex $\hat{S}{}^{\rm mat}_{cd}$ by
the new gravitational potential $\gamma^a$ and the
 one-loop two-point ``vertex''
${\mbox{\boldmath $\Gamma$}}_{cd}$, respectively:
        \begin{eqnarray}
        &&\gamma^a=-\frac 1{m_P^2} D^{ab}
        {\mbox{\boldmath $\Gamma$}}_b,\,\,\,
        {\mbox{\boldmath $\Gamma$}}_b\equiv
        \frac{\delta{\mbox{\boldmath $\Gamma$}}}{\delta g^b}, \label{6.4}\\
        &&{\mbox{\boldmath $\Gamma$}}_{cd}\equiv
        \frac{\delta^2{\mbox{\boldmath $\Gamma$}}}
        {\delta g^c\delta g^d}\ .                            \label{6.5}
        \end{eqnarray}
The new gravitational potential is generated by the 
one-loop stress tensor
of vacuum gravitons ${\mbox{\boldmath $\Gamma$}}_b$ which, similarly to
(\ref{5.11}), enters as a matter source in the linearized Einstein equations
for $\gamma^a$. As is known \cite{DW:Dynamical} (see Appendix A), this
vacuum stress is also covariantly conserved on shell,
        \begin{eqnarray}
        R^a_\mu{\mbox{\boldmath $\Gamma$}}_a
        =O(\delta S/\delta g)                              \label{6.6}
        \end{eqnarray}
(which implies on-shell gauge invariance of the effective action), whence it
follows that the new gravitational potential also satisfies the linearized
gauge condition
        \begin{eqnarray}
        \chi^\mu_a \gamma^a=0.
        \end{eqnarray}

Now the calculation of the cross term in $\hat{H}{}^{\rm mat}_1$ presents no
difficulty and gives
        \begin{eqnarray}
        -\frac 1{m_P^2} {\cal G}^{mn}
        ({\cal D}_m{\mbox{\boldmath$P$}}){\mbox{\boldmath$P$}}^{-1}\,
        ({\cal D}_n\hat{\mbox{\boldmath$U$}}_0)
        \hat{\mbox{\boldmath$U$}}{}^{-1}_0=
        m_P^2\,{\cal G}^{mn}\,
        (\stackrel{\rightarrow}{W}_{ma}\!\gamma^a)
        (\stackrel{\rightarrow}{W}_{nb}\!\hat{t}{}^b).      \label{6.7}
        \end{eqnarray}
The graphical representation of (7.4) is similar to Fig.~1,
with $\hat{t}{}^a$ ($\hat{S}{}^{mat}_{cd}$) replaced by
$\gamma^a$ ($\Gamma_{cd}$), and no time ordering in the first
graph. The graphical representation of (7.9) is very similar
to the first graph in Fig.~1, with {\em one} $\hat{t}{}^a$
replaced by $\gamma^a$.

\section{Effective Hamiltonian and back reaction}
\hspace{\parindent}
Collecting the equations (\ref{5.18}), (\ref{6.3}) and (\ref{6.7}) we
get the total effective Hamiltonian in the first order
 of the $1/m_P^2$-expansion
        \begin{eqnarray}
        &&\hat{H}{}^{\rm eff}_1=\hat{H}{}^{\rm mat}+
        \frac 12 m_P^2\,{\cal G}^{mn}\,{\rm T}
        \left(\stackrel{\rightarrow}{W}_{ma}\!\hat{h}{}^a\,
        \stackrel{\rightarrow}{W}_{nb}\!\hat{h}{}^b\right)
        +\frac{\I}{2} D^{ab}w_{abc}(t_+)\,\hat{h}{}^c\nonumber\\
        &&\qquad\qquad\qquad-\frac{\I}{2} {\cal G}^{mn}\,
        (\stackrel{\rightarrow}{W}_{ma}\!D^{ac})\,
        (\stackrel{\rightarrow}{W}_{nb}\!D^{bd})\,[\,S_{cde}\,
        \hat{h}{}^e
        +\frac 1{m_P^2}(\hat{S}{}^{\rm mat}_{cd}+
        {\mbox{\boldmath $\Gamma$}}_{cd})\,].     \label{6.8}
        \end{eqnarray}
Here the {\em full} gravitational potential
 $\hat{h}{}^a=\hat{t}{}^a+\gamma^a$
determines the total back reaction of quantum matter and graviton vacuum
polarization on the gravitational background,
        \begin{eqnarray}
        \hat{h}{}^a=-\frac 1{m_P^2} D^{ab}\,(\,\hat{T}_b+
        {\mbox{\boldmath $\Gamma$}}_b\,).              \label{6.9}
        \end{eqnarray}
It solves the linearized Einstein equations with the full stress tensor
source and satisfies the gauge and boundary conditions of the above type,
        \begin{eqnarray}
        &&m_P^2\,S_{ab}\hat{h}{}^b+
        \hat{T}_a+{\mbox{\boldmath $\Gamma$}}_a=0,      \label{6.10}\\
        &&\chi^\mu_a\hat{h}{}^a=0,\quad
        \hat{h}{}^i(t_\pm)=0.                            \label{6.11}
        \end{eqnarray}

This result was obtained in the first subleading order
 of the $1/m_P^2$-expansion,
but the different terms in (\ref{6.8}) have a very different nature from the
viewpoint of a physically reasonable approximation scheme. As it has already
been mentioned above, only the first two terms on the right-hand side have
tree-level contributions, while the rest are essentially loop
contributions. Formally
this is explicitly indicated by an extra inverse power of $m_P^2$. Certainly
this is an artifact of the definition of the gravitational potential
(\ref{6.9}) involving the inverse of $m_P^2$ and, therefore, formally all the
terms belong to the same order of
 asymptotic expansion in Planck mass. However,
from a physical point of view, one might consider quantum states with a big
mean value of matter energy density, much higher than the energy of graviton
vacuum polarization, so that
        \begin{eqnarray}
        \frac{\Big<\hat{T}_a\Big>}{m_P^2}=O(1),\,\,\,
        \frac{{\mbox{\boldmath $\Gamma$}}_a}{m_P^2}
        =O(1/m_P^2).                                    \label{6.12}
        \end{eqnarray}
In this situation, then, only the back reaction from the matter part
is relevant. This is, in fact, the situation most frequently
studied in the quantum field theory on curved backgrounds.
In this case of large matter sources,
$\hat{t}{}^a=O(1)\gg\gamma^a$, only the first two terms of
$\hat{H}{}^{\rm eff}_1$ remain dominating, and (\ref{6.8}) reduces to
        \begin{eqnarray}
        \hat{H}{}^{\rm eff}_1=\hat{H}{}^{\rm mat}+
        \frac 12 m_P^2\,{\cal G}^{mn}\,{\rm T}
        \left(\stackrel{\rightarrow}{W}_{ma}\!\hat{t}{}^a\,
        \stackrel{\rightarrow}{W}_{nb}\!\hat{t}{}^b\right)
        +O(1/m_P^2).                                    \label{6.13}
        \end{eqnarray}
Here we mainly restrict ourselves to this physical situation when the matter
field back reaction is dominating over other effects.

What is the interpretation of the second term in (8.6)? According to the
definition of the Wronskian operator (\ref{4.19}),
$\stackrel{\rightarrow}{W}_{ma}\!\hat{t}{}^a$ is the linearized momentum
conjugate to the gravitational potential $\hat{t}{}^a$, so that this term is
just the kinetic energy of the gravitational radiation produced due to the
back reaction of quantum matter sources.

 There is another important
representation of this term which allows one to understand better its
dynamical properties and establish its relation to an
analogous term found in \cite{Kiefer1}.
This representation is based on a special decomposition of the back reaction
momentum $\stackrel{\rightarrow}{W}_{ma}\!\hat{t}{}^a$ in the basis reflecting
the gauge properties of the gravitational field. We know that the gauge
direction in configuration space
 is defined by the generator $\nabla^i_\mu$ (\ref{42}).
By using the contravariant metric ${\cal G}^{mn}$ (\ref{3.9a}) and its
covariant inverse,
        \begin{eqnarray}
        {\cal G}_{mn}=(\,{\cal G}^{-1})_{mn}\ ,               \label{6.14}
        \end{eqnarray}
one can define in configuration space
 the orthogonal decomposition of tangent and cotangent vector
spaces into subspaces longitudinal and transverse to the gauge generator. The
projector onto the transverse subspace looks as follows. Define the matrix
        \begin{eqnarray}
        N_{\mu\nu}=\nabla^i_\mu {\cal G}_{ik} \nabla^k_\nu   \label{6.15}
        \end{eqnarray}
and assume that on the background of the classical extremal this matrix is
nondegenerate, denoting the inverse by
        \begin{eqnarray}
        N^{\mu\nu}=(\,N^{-1})^{\mu\nu}.                \label{6.16}
        \end{eqnarray}
Then the transverse projector equals
        \begin{eqnarray}
        &&\Pi^m_n=\delta^m_n-\nabla^m_\mu
        N^{\mu\nu}\nabla_{\nu n},                         \label{6.17}\\
        &&\nabla_{\nu n}\equiv\nabla_\nu ^m {\cal G}_{mn}. \label{6.18}
        \end{eqnarray}

The distinguished role of this decomposition (with respect to such a metric)
follows from the fact that the matrix (\ref{6.15}) defines the matrix of
second functional derivatives of the action with respect to Lagrange
multipliers -- lapse and shift functions
        \begin{eqnarray}
        \frac{\delta^2 S}{\delta N^\mu(t)\delta N^\nu(t')}=
        N_{\mu\nu}\delta(t-t')                              \label{6.19}
        \end{eqnarray}
and turns out to be the coefficient of the variation of these functions
in the linearized gravitational constraints $H_\mu(q,p^0(q,\dot{q},N))$.
This property is used in Appendix~C to show that the momentum of the
gravitational potential $\stackrel{\rightarrow}{W}_{ma}\!\hat{t}{}^a(t_+)$
taken at the final moment of time $t_+$ has the following orthogonal
decomposition
        \begin{eqnarray}
        \stackrel{\rightarrow}{W}_{ma}\!\hat{t}{}^a(t_+)=
        -\Pi_{mn}\dot{\hat{t}}{}^n(t_+)
        -\frac 1{m_P^2}\,\nabla_{m\mu}
        N^{\mu\nu}\hat{T}_\nu(t_+),                       \label{6.20}
        \end{eqnarray}
where $\Pi_{mn}={\cal G}_{mi}\Pi^i_n$. In view of this decomposition
the effective Hamiltonian acquires the form
        \begin{eqnarray}
        \hat{H}{}^{\rm eff}_1=\hat{H}{}^
{\rm mat}+
        \frac 12 m_P^2\,\Pi_{mn}\,{\rm T}
        \left(\,\dot{\hat{t}}{}^m\,\dot{\hat{t}}{}^n\right)
        +\frac 1{2m_P^2}\,N^{\mu\nu}\hat{T}_\mu(t_+)
        \hat{T}_\nu(t_+)+O(1/m_P^2)                       \label{6.21}
        \end{eqnarray}
(this should not look confusing because $\hat{T}_\mu(t_+)/m_P^2=O(1)$ and the
third term is therefore dominating over $O(1/m_P^2)$).

An important difference between the second (kinetic) term
 and the third (potential)
term concerns their locality properties. The kinetic term is essentially
nonlocal, because in integral form it involves the fields at all moments of
time $t_+\geq t\geq t_-$. On the contrary, the third (potential) term is
ultralocal in all field variables taken at $t_+$. Moreover, $\mu$-components
of matter stress tensor coincide (up to sign)
with matter parts of the Hamiltonian and momentum constraints
$\hat{H}{}^{\rm mat}_\mu(t_+)$. In addition, at $t_+$ the Heisenberg
operators coincide with the Schr\"odinger ones
 (see the definition (\ref{5.3})).
Therefore, this part of the effective Hamiltonian is just a
quadratic combination, local in time,
 of the operators $\hat{H}{}^{\rm mat}_\mu$ taken
in the Schr\"odinger picture,
        \begin{eqnarray}
        \frac 1{2m_P^2}\,N^{\mu\nu}\hat{H}{}^{\rm mat}_\mu
        \hat{H}{}^{\rm mat}_\nu.                       \label{6.21a}
        \end{eqnarray}
This is exactly the structure in the effective iterational Hamiltonian
captured in \cite{Kiefer1}\footnote{Cf. the first correction term
in Eq.~(42) in that paper. This term does not contain
$\hbar$ explicitly, reflecting the tree-level nature of this term.
The remaining, imaginary, correction terms in Eq.~(42) of
\cite{Kiefer1} are proportional to $\hbar$
and are contained in the imaginary terms
of (8.1) in the present paper.}.
This term quadratic in the superhamiltonian of matter fields was
obtained there by calculating the ``longitudinal'' part of
$q_+$-derivatives of the above type. As we see, extension to the
case of the full configuration space
 and taking into account the ``transversal'' part
of these derivatives results in the local quadratic term (\ref{6.21a})
{\em plus} the nonlocal kinetic term
 involving only the transversal velocities of
gravitational potentials,
$\dot{\hat{t}}{}^m_\perp\equiv\Pi^m_n\dot{\hat{t}}{}^m$.

It should be emphasized that the representation (\ref{6.21}) and the
canonical decomposition (\ref{6.20}) exist only for problems with the
invertible operator (\ref{6.15}). This is an important restriction
characteristic of the problems related to the thin-sandwich problem in
which the nondegeneracy of (\ref{6.15}) guarantees the possibility to solve
the constraint equations for fixed superspace coordinates $q=g_{ab}({\bf x})$
and their velocities $\dot{q}=\dot{g}_{ab}({\bf x})$. This property is
violated, for example, for the linearized theory on flat-space background,
or more generally on backgrounds with Killing symmetries \cite{BarvU}.
For such backgrounds a similar decomposition should be modified by special
techniques in the kernel of the operator (\ref{6.15}) spanned by the
Killing vectors of the background \cite{BarvU}. This goes beyond the present
paper and will be considered elsewhere. In contrast to the representation
(\ref{6.21}), the original form of the effective Hamiltonian (\ref{6.13})
remains valid on all possible backgrounds.

\section{Summary and Outlook}
\hspace{\parindent}
The main result of our paper is the calculation of {\em all}
quantum gravitational correction terms to a given matter
Hamiltonian up to order $m_P^{-2}$, the final expression
being given in (8.1). This result both generalizes
previous results \cite{Kiefer1} and gives an interpretation
in terms of Feynman diagrams. Our discussion thus builds a bridge
between the canonical and the covariant frameworks in the
semiclassical approximation. (Since there is no spacetime at the
fundamental level, this is all one can do in this respect.)
In future applications we intend to apply these correction terms
to concrete physical situation such as the one in \cite{Kiefer},
where a quantum gravity-induced energy shift to the trace anomaly
in De~Sitter space was calculated.

There are, of course, still open issues. One important open point is
to find a consistent regularization that preserves the ordering
of \cite{BKr,Barv} without producing anomalies. This is a
contentious issue, and it was argued, for example, in \cite{CJZ}
that anomalies necessarily occur which may even spoil the standard
semiclassical approximation.

The perturbative non-renormalizability of quantum general relativity
can of course not be circumvented by the present approximation
scheme. Thus, either full canonical quantum gravity is consistent
and the approximation scheme if taken at all orders becomes useless,
or the full theory is inconsistent and one has to consider
an alternative approach such as superstring theory.
But even in the latter case it might be very likely that the
first quantum gravitational correction terms have the form
presented in (8.1). This hope is based on the correspondence principle
between the fundamental superstring theory and the low-energy classical
gravity theory and local theory of renormalizable gauge fields. The
progress of these theories teaches that there should exist a sub-Planckian
energy domain in which the predictions of {\em covariantly regularized
local} quantum gravity consistently describe the low-energy limit of
this fundamental theory of extended objects. Among the examples of
applications in such an energy domain one can mention the recently
proposed mechanism of generating the energy scale of inflation by
loop effects in quantum cosmology \cite{qcr} -- the phenomenon which,
being on one hand essentially quantum gravitational one, on the other hand
provides an effective suppression of sub-Planckian scales and thus
justifies the semiclassical expansion.

In our discussion we have considered states of the form (2.13),
i.e. states where the dominant part is a phase obeying
 the Einstein-Hamilton-Jacobi
equations. However, generally arbitrary
superpositions of such states are expected to occur. In such a
case there is no longer a unique background spacetime available
as the starting point for the approximation scheme.
In many realistic situations one can, however, understand how the
various semiclassical components in the superposition become
dynamically independent. The key mechanism is the process of
decoherence by irrelevant degrees of freedom \cite{dec}.
The results presented in our paper yield the necessary technical
prerequisites to study decoherence processes in the early universe
at higher orders of the inverse Planck mass.
It must also be mentioned that there might exist situations
where already the lowest order of the semiclassical approximation
breaks down \cite{LMO}. This will be considered elsewhere.

Finally, there is the question whether some of the quantum gravitational
correction terms lead to an effective violation of unitarity in the
matter sector. This is supported by the fact that imaginary terms
occur explicitly, see (8.1). It has previously been shown that
the occurrence of such terms would have important effects for the
process of black hole evaporation \cite{KMS}.
However, it was argued that one can perform the splitting of the
full wave function satisfying the Wheeler-DeWitt equation
into gravitational and matter parts in such a way that this
effective non-unitarity disappears \cite{Ital,Kim}.
We hope to clarify this issue in a future publication.

\renewcommand{\theequation}{\Alph{section}.\arabic{equation}}
\appendix
\section{Ward identities and properties of gravitational
potentials}
\hspace{\parindent}
We begin this Appendix by deriving the solution (\ref{4.24}) of the boundary
value problem (\ref{4.11}), (\ref{4.2}). For this purpose we write
first down the Wronskian relation
(\ref{4.18}) with $\varphi^a_1=\delta g^a(t)$ and $\varphi^b_2=D^{bc}(t,t')$,
integrated over $t$ from $t_-$ to $t_+$. On the
left-hand side we have $\delta g^c(t')$ in view of the equations for
$\delta g^a(t)$ and $D^{bc}(t,t')$, while the right-hand side gives only
the contributions of upper and lower limits of integration:
        \begin{eqnarray}
        \delta g^c(t')=
        \left.\left(\delta g^a \stackrel{\rightarrow}{W}{\!}^F_{ab}
        D^{bc}(t,t')-
        \delta g^a \stackrel{\leftarrow}{W}{\!}^F_{ab}D^{bc}(t,t')\right)
        \,\right|_{\,t=t_-}^{\,t=t_+}\ .
        \end{eqnarray}
In view of the boundary conditions on the Green function
(\ref{4.22}) -- (\ref{4.23})
and the relation (\ref{4.20}) for
the Wronskian operators of $F_{ab}$ and $S_{ab}$,
only one term $\delta q^i_+\stackrel{\rightarrow}{W}_{ib}
D^{bc}(t,t')$ survives on the right-hand side, because
$\stackrel{\rightarrow}{W}_{\mu b}=0$, and $\delta g^a$ by construction
satisfies the linearized gauge conditions (\ref{4.10}). This proves
(\ref{4.24}) and (\ref{4.25}).

Now let us prove that in (\ref{4.24}) $\delta g^a$ really
satisfies the linearized gauge conditions. For this purpose let us use
the Ward identity relating the gauge and ghost Green functions.
Eq. (\ref{4.13}) implies that
        \begin{eqnarray}
        c_{\alpha\beta}\stackrel{\rightarrow}{\chi}{\!}^\beta_b(d/dt)
        D^{bc}(t,t')=-Q^{-1\,\beta}_{\,\alpha}(t,t')
        \stackrel{\leftarrow}
        {R}{\!}^c_\beta(d/dt'),       \label{A.1}
        \end{eqnarray}
where $Q^{-1\,\beta}_{\,\alpha}(t,t')$ is a ghost Green function
subject to {\em Dirichlet} boundary conditions at $t_\pm$,
        \begin{eqnarray}
        Q^\alpha_\mu Q^{-1\,\beta}_{\,\alpha}=
        \delta^\beta_\mu,\,\,\,
        Q^{-1\,\beta}_{\,\alpha}(t_\pm,t')=0.     \label{A.2}
        \end{eqnarray}

Now acting on (\ref{4.24}) by $\chi^\mu_a$ and using (\ref{A.1}), one can
see that in view of the differential structure of the gauge generators
(\ref{4.4}) and (\ref{4.5}) and
the Wronskian operators, the time derivatives acting
on the ghost Green function at $t_+$ cancel and the rest of the terms vanish
due to the Dirichlet boundary conditions for $Q^{-1\,\beta}_{\,\alpha}$,
whence it follows that
        \begin{eqnarray}
        \chi^\mu_a\frac{\partial g^a}{\partial q^i_+}=0.
        \end{eqnarray}
The linearized gauge condition (\ref{4.10}) is thus fulfilled.
In proving this we used the expressions for Wronskian operators
        \begin{eqnarray}
        &&\stackrel{\rightarrow}{W}_{ik}(d/dt)=-{\cal G}_{ik}\frac d{dt}
        +{\rm local\,\,terms},         \label{A.3}  \\
        &&\stackrel{\rightarrow}{W}_{i\mu}=-\frac{\partial^2 L(q,\dot{q},N)}
        {\partial\dot{q}{}^i\,\partial N^\mu}=
        \frac{\partial H_\mu(q,p^0(q,\dot{q},N))}{\partial\dot{q}{}^i}
        ={\cal G}_{ik}\nabla^k_\mu\ .    \label{A.4}
        \end{eqnarray}

Let us now turn to the gauge property of the gravitational potentials
$\hat{t}{}^a$ and $\gamma^a$. Acting on (\ref{5.10}) by the gauge matrix
and using the Ward
identity of the above type, one can see that the result is proportional to
$R^a_\mu\hat{T}_a$ which is zero in view of (\ref{5.13}). This proves
(\ref{5.12}). Similar proof holds for the gravitational potential
$\gamma^a$.

\section{The vertex $w_{abc}(t_+)$}
\hspace{\parindent}
The derivation of the vertex begins by noting that in view of
(\ref{A.3}) and (\ref{A.4})
        \begin{eqnarray}
        \stackrel{\rightarrow}{W}_{ma}\!\hat{t}{}^a(t_+)=
        -{\cal G}_{mn}\frac{d\hat{t}{}^n(t_+)}{dt_+}+
        \nabla_{m\mu}\hat{t}{}^\mu(t_+)\ .          \label{C.1}
        \end{eqnarray}
Therefore, the calculation of the derivative on the right-hand side of
(\ref{5.17}) in terms of the background field variation,
$\partial/\partial q_+=(\partial g^a/\partial q_+)\delta/\delta g^a$,
takes the form
        \begin{eqnarray}
        ({\cal G}^{mn}{\cal D}_m\!\stackrel{\rightarrow}{W}_{ma})
        \,\hat{t}{}^a\,&=&\,
        -\stackrel{\rightarrow}{W}_{ma}D^{a\perp}(t_+,t_+)\,
        {\cal G}^{mn}_\perp\,\stackrel{\rightarrow}{W}_{nb}
        \hat{t}{}^b(t_+)                                   \nonumber\\
        &&+\stackrel{\longrightarrow}{(\delta R^m_\mu/\delta g^b)}
        D^{bc}(t_+,t_+)
        \stackrel{\leftarrow}{W}_{cm}\,\hat{t}{}^\mu(t_+),  \label{B.1}
        \end{eqnarray}
where the variation of $\nabla_{m\mu}$ in (\ref{C.1}) is replaced by
the variation of the Lagrangian generator
$R^m_\mu=\nabla^m_\mu|_{p=p^0}$, and it was taken into account
that the covariant superspace derivative of the three-entry object
${\cal G}^{mn}_\perp$ is zero. The kernel of the first-order differential
operator $\stackrel{\longrightarrow}{(\delta R^m_\mu/\delta g^b)}$ is
understood here as acting on the first argument of the Green function.
Thus this expression really reduces to the one-loop tadpole structure --
the coincidence limit of the graviton Green function and its derivatives
-- and it can be represented by a quasi-local vertex $w_{abc}(t_+)$
read off (\ref{B.1}).

\section{Gauge decomposition of gravitational potentials}
\hspace{\parindent}
The proof of the gauge decomposition of the momentum conjugated to the
gravitational potential (\ref{6.20}) is based on (\ref{C.1}) above.
The $\mu$-component of the gravitational potential (\ref{5.10}) there
involves
the $D^{\mu a}(t_+,t)$-component of the Green function. It can be found
from the $\mu$-component of (\ref{4.21})
        \begin{eqnarray}
        \stackrel{\rightarrow}{S}_{\mu i}\!D^{ia}(t_+,t)
        +\stackrel{\rightarrow}{S}_{\mu\nu}\!D^{\nu a}(t_+,t)
        =\stackrel{\rightarrow}{\chi}{\!}^\alpha_\mu c_{\alpha\beta}\!
        \stackrel{\rightarrow}{\chi}{\!}^\beta_b D^{ba}(t_+,t).  \label{C.2}
        \end{eqnarray}
Let us use the Ward identity (\ref{A.1}) and take into account the Dirichlet
boundary condition on the ghost Green function. The latter implies that
on the right-hand side here only the time derivative part of
$\stackrel{\rightarrow}{\chi}{}^\alpha_\mu=
-(\partial\chi^\alpha/\partial\dot{N}{}^\mu)d/dt+\ldots$ will survive acting on
the ghost propagator at $t_+$. Similarly, on the left-hand side the operator
coefficients $\stackrel{\rightarrow}{S}_{\mu i}$ and
$\stackrel{\rightarrow}{S}_{\mu\nu}$ are the first-order
differential operator and ultralocal operator given by
        \begin{eqnarray}
        \stackrel{\rightarrow}{S}_{\mu i}=
        -\frac{\partial^2 L(q,\dot{q},N)}
        {\partial\dot{q}{}^i\,\,\partial N^\mu}\frac d{dt}+
        {\rm local\,terms}=-\nabla_{\mu i}\frac d{dt}      \label{C.3}
        +{\rm local\,terms}
        \end{eqnarray}
and (\ref{6.19}), respectively. According to our assumption, the
functional matrix $N_{\mu\nu}$ in (\ref{6.19}) is invertible, so that the
equation resulting from (\ref{C.2}) can be solved with respect to
$D^{\nu a}(t_+,t)$,
        \begin{eqnarray}
        D^{\mu a}(t_+,t)=N^{\mu\nu}\nabla_{\mu m}\frac d{dt_+}
        D^{ma}(t_+,t)+N^{\mu\nu}\frac{\partial\chi^\alpha}
        {\partial\dot{N}{}^\nu}\frac d{dt_+}
        Q^{-1 \beta}_{\,\alpha}(t_+,t)
        \stackrel{\leftarrow}{R}{}^a_\beta(d/dt).          \label{C.4}
        \end{eqnarray}

Note that the $a=\sigma$-component of the generator $R^a_\beta$
($\stackrel{\leftarrow}{R}{}^\sigma_\beta
=\stackrel{\leftarrow}{d/dt}\delta^\sigma_\beta+\dots$) is acting to the left
in this equation. When contracting it with $\hat{T}_a$ to obtain
$\hat{t}{}^\mu(t_+)$
one should integrate in the last term by parts in order to use the
conservation of matter stress
$\stackrel{\rightarrow}{R}{}^a_\beta\hat{T}_a=0$ which yields
        \begin{eqnarray}
        \hat{t}{}^\mu(t_+)=N^{\mu\nu}\nabla_{\mu m}\frac
        {d\hat{t}{}^m}{dt_+}-\left.\frac 1{m_P^2}\,
        N^{\mu\nu}\frac{\partial\chi^\alpha}
        {\partial\dot{N}{}^\nu}\frac d{dt_+}
        Q^{-1 \beta}_{\,\alpha}(t_+,t)
        \hat{T}_\beta(t)\,\right|_{t=t_+}.                          \label{C.5}
        \end{eqnarray}
The contribution of the surface term at $t_+$ here is not zero even for
the Dirichlet boundary conditions, because the
coincidence limit at $t_+$ is understood in the sense when the second argument
of $(d/dt_+)Q^{-1 \beta}_{\,\alpha}(t_+,t')$ tends to $t_+$ {\em after}
the first one. It is easy to show that
        \begin{eqnarray}
        \frac{\partial\chi^\alpha}
        {\partial\dot{N}{}^\mu}\frac d{dt_+}
        \left.Q^{-1 \beta}_{\,\alpha}(t_+,t)\,\right|_{t=t_+}=
        \left.\stackrel{\rightarrow}{W}{\!}_\mu^{\,\,\,\alpha}(d/dt_+)
        Q^{-1 \beta}_{\,\alpha}(t_+,t)\,\right|_{t=t_+}
        =\delta_\mu^\beta\,
        \end{eqnarray}
where $\stackrel{\rightarrow}{W}{}_\mu^{\,\,\,\alpha}(d/dt_+)$ is a
Wronskian operator corresponding to the Faddeev-Popov operator (\ref{4.9}).
The second equality here can be easily proved from the spectral
decomposition of the ghost Green function subject to Dirichlet boundary
conditions\footnote
{
Note that this property is a direct analogue of the following
equality for the matrix (\ref{4.25}):
$\partial q^k(t_+)/\partial q^i_+=\delta^k_i$.
}
\cite{reduct}.

Thus
        \begin{eqnarray}
        \hat{t}{}^\mu(t_+)=N^{\mu\nu}\nabla_{\nu m}\frac
        {d\hat{t}{}^m(t_+)}{dt_+}-\frac 1{m_P^2}\,
        N^{\mu\nu}\hat{T}_\nu(t_+).
        \end{eqnarray}
Substituting this relation into (\ref{C.1}) we finally get the decomposition
(\ref{6.20}). It is obvious that the same proof holds for the gauge
decomposition of the one-loop gravitational potential $\gamma^a$.

\vskip 5mm
\begin{center}
{\bf Acknowledgements}
\end{center}
A.O.B. is deeply grateful for the warm hospitality at the Department
of Physics of the University of Freiburg where a major part of this work has
been done due to the financial support by the
 DFG-grant 436 RUS 113/333/2. The work of A.O.B. was
also supported by the Russian Foundation for Basic Research
under grants 96-02-16287 and 96-02-16295 and the European
Community Grant INTAS-93-493-ext. Partly this work has been also made
possible due to the support by the Russian Research Project
``Cosmomicrophysics''.

\newpage
\begin{center}
{\bf Figure Caption}
\end{center}
{\bf Fig 1}. Contributions of quantum matter fields to $\hat{H}^{\rm eff}_1$,
see (6.20). The time parameter
$t_+$ labels the vertices at the spacetime point at which
$\hat{H}^{\rm eff}_1$ is evaluated.
 Dashed lines labelled by $\mbox{T}$ denote the chronological
ordering of matter stress tensors in the bilinear combinations of
gravitational potentials $\hat{t}^a$.

\end{document}